\documentclass[a4paper,fleqn,usenatbib]{mnras}

\usepackage[T1]{fontenc}
\usepackage{ae,aecompl}

\usepackage{graphicx}	
\usepackage{amsmath}	
\usepackage{amssymb}	
\usepackage[compatibility=false]{caption}
\usepackage{subcaption}
\usepackage{cleveref}
\usepackage{xspace}
\usepackage[super]{nth}
\usepackage{multirow}

\usepackage{etoolbox}
\makeatletter
\patchcmd\@combinedblfloats{\box\@outputbox}{\unvbox\@outputbox}{}{%
   \errmessage{\noexpand\@combinedblfloats could not be patched}%
}%
\makeatother

\usepackage{tikz}
\usetikzlibrary{bayesnet}

\graphicspath{{plots/}}

\crefname{figure}{Fig.}{Figs.}
\crefname{equation}{Eq.}{Eqs.}
\crefname{section}{\S}{\S\S}
\crefname{chapter}{\S}{\S\S}
\crefname{table}{Table}{Tables}

\usepackage{lipsum}
\usepackage{xifthen}
\newcommand{\lorem}[1][]{
    \ifthenelse{\isempty{#1}}
      {\def\loremStep{1}}
      {\def\loremStep{#1}}
    \setcounter{loremB}{\theloremA}
    \addtocounter{loremB}{\loremStep}
    \addtocounter{loremB}{-1}
    \lipsum[\theloremA-\theloremB]
    \addtocounter{loremA}{\loremStep}
}
\newcounter{loremA}
\newcounter{loremB}

\newcommand{\gps}{\ensuremath{g_{\mathrm{P1}}}\xspace}
\newcommand{\rps}{\ensuremath{r_{\mathrm{P1}}}\xspace}
\newcommand{\ips}{\ensuremath{i_{\mathrm{P1}}}\xspace}
\newcommand{\zps}{\ensuremath{z_{\mathrm{P1}}}\xspace}
\newcommand{\yps}{\ensuremath{y_{\mathrm{P1}}}\xspace}
\newcommand{\Eab}[2]{\ensuremath{\mathrm{E} \! \left( {#1} \! - \! {#2} \right)}\xspace}
\newcommand{\Erz}{\Eab{\rps}{\zps}}
\newcommand{\EBV}{\Eab{B}{V}}
\renewcommand{\deg}[1]{\ensuremath{{#1}^{\circ}}\xspace}

\newcommand{\vect}[1]{\ensuremath{\mathbf{#1}}}
\renewcommand{\arg}[1]{\ensuremath{\! \left( #1 \right)}}

\title[3D Galactic Reddening, v2.0]{Galactic Reddening in 3D from Stellar Photometry -- An Improved Map}

\author[G. M. Green et al.]
{Gregory M. Green,$^{1,2}$\thanks{E-mail: gregorymgreen@gmail.com}
Edward F. Schlafly,$^{3,4}$
Douglas Finkbeiner,$^{5}$
\newauthor
Hans-Walter Rix,$^{6}$
Nicolas Martin,$^{6,7}$
William Burgett,$^{8}$
Peter W. Draper,$^{9}$
\newauthor
Heather Flewelling,$^{10}$
Klaus Hodapp,$^{10}$
Nicholas Kaiser,$^{10}$
\newauthor
Rolf-Peter Kudritzki,$^{10}$
Eugene A. Magnier,$^{10}$
Nigel Metcalfe,$^{9}$
\newauthor
John L. Tonry,$^{10}$
Richard Wainscoat,$^{10}$
Christopher Waters$^{10}$
\\
$^{1}$Kavli Institute for Particle Astrophysics and Cosmology, Physics and Astrophysics Building, 452 Lomita Mall, Stanford, CA 94305, USA\\
$^{2}$Porat Fellow\\
$^{3}$Lawrence Berkeley National Laboratory, One Cyclotron Road, Berkeley, CA 94720, USA\\
$^{4}$Hubble Fellow\\
$^{5}$Harvard Astronomy, Harvard-Smithsonian Center for Astrophysics, 60 Garden St., Cambridge, MA 02138, USA\\
$^{6}$Max-Planck-Institut f\"ur Astronomie, K\"onigstuhl 17, D-69117 Heidelberg, Germany\\
$^{7}$Universit\'e de Strasbourg, CNRS, Observatoire astronomique de Strasbourg, UMR 7550, F-67000 Strasbourg, France\\
$^{8}$GMTO Corporation, 465 N. Halstead Street, Suite 250, Pasadena, CA 91107, USA\\
$^{9}$Durham University, Department of Physics, Science Laboratories, South Road, Durham DH1 3LE, UK\\
$^{10}$University of Hawaii, Institute for Astronomy, 2680 Woodlawn Dr., Honolulu, HI 96822, USA
}

\date{Accepted XXX. Received YYY; in original form ZZZ}

\pubyear{2018}

\begin{document}
\label{firstpage}
\pagerange{\pageref{firstpage}--\pageref{lastpage}}
\maketitle

\begin{abstract}
We present a new 3D map of interstellar dust reddening, covering three quarters of the sky (declinations of $\delta \gtrsim -\deg{30}$) out to a distance of several kiloparsecs. The map is based on high-quality stellar photometry of 800 million stars from Pan-STARRS~1 and 2MASS. We divide the sky into sightlines containing a few hundred stars each, and then infer stellar distances and types, along with the line-of-sight dust distribution. Our new map incorporates a more accurate average extinction law and an additional 1.5 years of Pan-STARRS~1 data, tracing dust to greater extinctions and at higher angular resolutions than our previous map. Out of the plane of the Galaxy, our map agrees well with 2D reddening maps derived from far-infrared dust emission. After accounting for a 15\% difference in scale, we find a mean scatter of $\sim$10\% between our map and the Planck far-infrared emission-based dust map, out to a depth of $0.8 \, \mathrm{mag}$ in \Erz, with the level of agreement varying over the sky. Our map can be downloaded at \url{http://argonaut.skymaps.info}, or by its DOI: \href{http://dx.doi.org/10.7910/DVN/LCYHJG}{10.7910/DVN/LCYHJG}.
\end{abstract}




\section{Introduction}
\label{sec:introduction}

Interstellar dust provides foregrounds for a wide range of astrophysical observations. At ultraviolet (UV), optical and near-infrared (NIR) wavelengths, scattering and absorption by dust grains can cause significant, or even complete extinction. Because this scattering and absorption is wavelength-dependent, dust both dims and reddens background sources. Any UV, optical or NIR observation that relies either on a precise magnitude calibration, such as measurements of the SNe Ia redshift--luminosity distance relation, or on a precise color calibration requires an accurate accounting for the effects of dust extinction. At longer wavelengths, in the far-infrared (FIR) and sub-millimeter regions of the spectrum, $\sim \! 20 \, \mathrm{K}$ dust radiates strongly. For Cosmic Microwave Background (CMB) cosmology, foreground dust emission must be disentangled from the background cosmological signal. Finally, at frequencies of 30--300 GHz ($\lambda \approx 1$--10~mm) polarized dust emission is a dominant foreground in searches for B-mode polarization in the CMB. Thus, over a wide swath of the electromagnetic spectrum, dust extinction and emission pose significant challenges.

Yet dust is not only a nuisance, but is itself an important component of the structure of our Milky Way Galaxy. Dust traces the interstellar medium, and provides one means of exploring the structure and dynamics of the Galaxy. Dust also plays a critical role in the chemical dynamics of the interstellar medium, depleting it of some elements and catalyzing the formation of molecular hydrogen. Dust also plays an important role in star formation, while providing us a means of tracing star-forming regions of the Galaxy.

Until recently, the standard dust maps in use have traced the distribution of dust only in two dimensions -- that is, with Galactic latitude and longitude, neglecting the distribution of dust with distance. \citet[``BH'']{Burstein1978a,Burstein1978b,Burstein1982} developed the first modern map of dust reddening across most of the sky, using \textsc{Hi} column density and galaxy counts as proxies for dust reddening. In the past two decades, \citet[hereafter ``SFD'']{Schlegel1998} has been a standard reference for Galactic reddening. SFD fit a model of dust temperature and optical depth to IRAS 100~$\mu$m and DIRBE 100~$\mu$m and 240~$\mu$m all-sky maps, and then calibrated optical depth at 100~$\mu$m to dust reddening using a sample of elliptical galaxies with well-determined reddenings. \citet{Schlafly2011} recalibrated this relation using the blue tip of the stellar locus. \citet[``Planck14'']{PlanckCollaboration2013} applied a similar method as SFD to far-infrared Planck and IRAS maps.

In the past decade, work has progressed on mapping dust in 3D. \citet{Marshall2006a} computed a three-dimensional map of dust reddening, using post-main sequence stars observed by the Two Micron All-Sky Survey \citep[``2MASS'']{Skrutskie2006} as tracers of the dust. \citet{Berry2011} used maximum-likelihood estimates of stellar reddening and distance, based on Sloan Digital Sky Survey \citep[``SDSS'']{York2000} photometry, to trace dust reddening in three dimensions.

More recently, work has turned towards probabilistic methods that infer stellar properties and the distribution of dust in a self-consistent, Bayesian framework. \citet[``Sale14'']{Sale2014b} maps extinction across the northern Galactic plane, applying a hierarchical model of stellar parameters and line-of-sight dust reddening to photometry from the INT Photometric H$\alpha$ Survey of the Northern Galactic Plane \citep[``IPHAS'']{Drew2005-IPHAS}. \citet{Chen2014} uses optical photometry from the Xuyi Schmidt Telescope Photometric Survey of the Galactic Anti-center (XSTPS-GAC), NIR photometry from 2MASS and MIR photometry from the Wide-field Infrared Survey Explorer \citep[``WISE'']{Wright2010} to produce a 3D map of reddening towards the Galactic anticenter. \citet{Lallement2013} computes a 3D map of local interstellar reddening using stellar parallaxes and reddening estimates, with the added assumption that the density of dust is spatially smooth.

The present paper builds on the work of \citet{Green2014-method}, which lays out a method for inferring the distribution of dust reddening in three dimensions, and \citet[hereafter, ``Bayestar15'']{Green2015-release}, which applies this method to Pan-STARRS~1 \citep[``PS1'']{Chambers2016-PS1} and 2MASS photometry to produce a 3D map of dust reddening across three quarters of the sky. Since the publication of Bayestar15, PS1 has completed its 5-band optical/NIR survey of three quarters of the sky, providing more homogeneous and deeper coverage than was previously available. This paper makes use of the newer PS1 photometry in order to present an updated map of dust reddening, covering the same three quarters of the sky, corresponding to $\delta \gtrsim -30^{\circ}$. This paper also makes use of the empirical mean reddening law found by \citet{Schlafly2016-extcurve}, which provides a better description of extinction of stellar light due to dust in the PS1 and 2MASS passbands than the reddening law used by Bayestar15. Finally, this paper adopts a somewhat different strategy for determining what angular resolution to use in different regions of the sky. We refer to the new map presented in this paper as ``Bayestar17.''

The paper is organized as follows. \cref{sec:method} reviews our method of inferring the 3D distribution of dust reddening, and discusses changes made to our model since the publication of Bayestar15. Next, \cref{sec:data} describes the PS1 and 2MASS photometry which we make use of. \cref{sec:results} presents Bayestar17 and compares it with previous reddening maps. Finally, \cref{sec:discussion} discusses advantages and disadvantages of our mapping technique, suggests possible extensions to the method, and discusses future datasets which will be applied to 3D dust mapping.

\begin{figure*}
    \begin{center}
        \begin{tikzpicture}[x=2.2cm,y=1.8cm]
            \node[latent, label={type}]
                  (Theta)
                  {$\Theta$} ;
            \node[det, right=of Theta, xshift=0cm,
                  label={[label distance=-2pt]90:abs. mag.}]
                  (M)
                  {$\vect{M}$} ;
            \node[det, right=of M, xshift=0cm,
                  label={[label distance=-2pt, xshift=-5pt]90:apparent mag.}]
                  (m)
                  {$\vect{m}$} ;
            \node[det, below=of m,
                  label={[yshift=1pt]0:extinction}]
                  (A)
                  {$\vect{A}$} ;
            \node[latent, left=of A, yshift=1cm,
                  label={180:distance modulus}]
                  (mu) {$\mu$} ;
            \node[latent, left=of A, yshift=-0.25cm,
                  label={180:extinction offset}]
                  (delta)
                  {$\delta$} ;

            \factor[right=of m, xshift=0.25cm]
                    {m-factor}
                    {$\mathcal{N}$}
                    {} {} ;

            \node[obs, right=of m-factor, xshift=0cm,
                  label={[label distance=0pt, xshift=-8pt]-90:observed mag.}]
                  (m-obs)
                  {$\vect{m}_{\mathrm{obs}}$} ;
            \node[const, below=of m-factor, yshift=1cm,
                  label={[label distance=0pt]-90:obs. uncertainty}]
                  (var-m)
                  {$\Sigma_{m}$} ;

            \node[det, below=of A, yshift=0.5cm,
                  label={0:line-of-sight extinction}]
                  (A-mu)
                  {$\vect{A} \arg{\mu}$} ;

            \node[latent, left=of A-mu, xshift=0.5cm,
                  label={180:l.o.s. dust distribution}]
                  (alpha)
                  {$\vect{\alpha}$} ;

            \node[const, below=of A-mu, yshift=0.20cm,
                  label={180:extinction vector}]
                  (R)
                  {$\vect{R}$} ;

            \edge {Theta} {M} ;
            \edge {M,mu,A} {m} ;
            \edge {alpha,R} {A-mu} ;
            \edge {mu,A-mu,delta} {A} ;

            \factoredge {m,var-m} {m-factor} {m-obs} ;

            \plate {star} {(Theta) (M) (mu) (A) (m-factor) (m-obs) (var-m)} {$i = 1 , \ldots , n_{\mathrm{stars}}$} ;
            \plate {distance} {(alpha)} {$\hspace{1.3cm} k = 1 , \ldots , n_{\mathrm{distances}}$} ;
            \plate {sightline} {(star) (distance)} {$j = 1 , \ldots , n_{\mathrm{pixels}}$} ;
        \end{tikzpicture}
    \end{center}
    \caption{Directed factor graph \citep{Dietz2010} of the Bayestar model. Unfilled circles represent parameters to be inferred, diamonds represent parameters that are deterministic functions of their inputs, and shaded circles represent observed variables. Constants (such as $\vect{R}$) are not enclosed, while solid squares represent random processes. In the above diagram, the solid square labeled $\mathcal{N}$ represents the measurement process, which introduces Gaussian noise into the observed apparent magnitudes.}
    \label{fig:bayestar-plate-diagram}
\end{figure*}
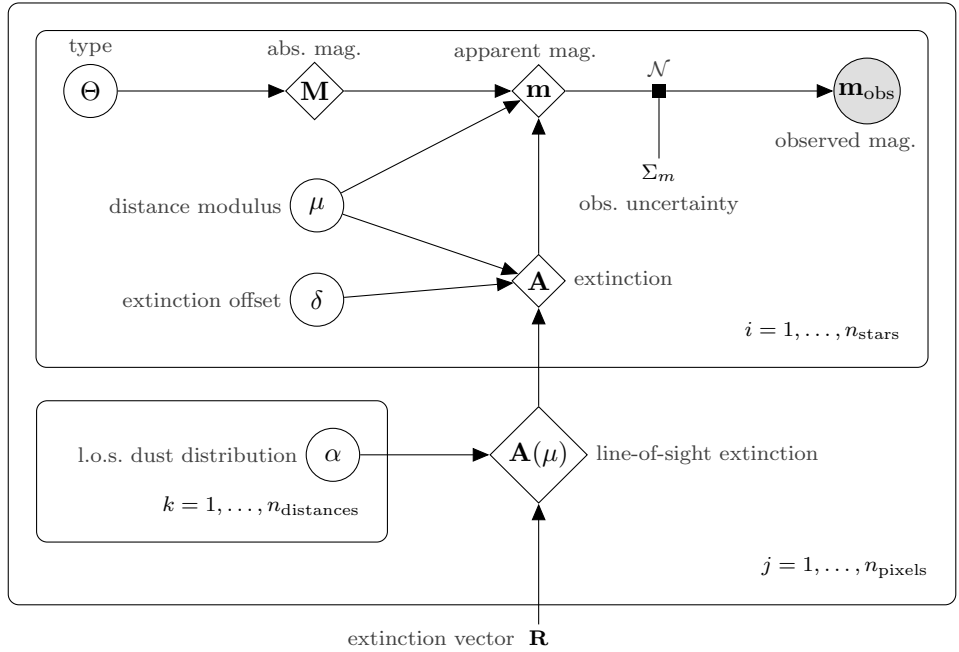

\section{Method}
\label{sec:method}

Our overall goal is to determine the dust extinction throughout the Galaxy, as a function of both angular position on the sky and distance from the Solar System. We use stars as tracers of the dust column. Interstellar dust both reddens and dims stellar light. If all stars were of a standard color and luminosity, the task of determining the three-dimensional distribution of dust would be trivial, since it would be immediately possible to determine both the distance and foreground dust column of any star. However, stars have different intrinsic colors and luminosities. When we try to infer stellar distance and foreground dust column, we have to marginalize over the full range of possible stellar properties (e.g., masses, ages and metallicities). As a consequence, when we observe the broadband photometry of a star, we can only probabilistically determine its distance and foreground dust column.

We group stars into small angular patches -- sightlines. Based on broadband photometric measurements of each star, we compute a probability distribution over the star's distance and foreground dust column. Each star puts a constraint on the line-of-sight distance vs. dust column relation. With hundreds of stars along a single sightline, we are able to put a strong constraint on the dust density as a function of distance. We assume that the reddening law is constant throughout the Galaxy.

The method used here largely follows that of \citet{Green2014-method,Green2015-release}, with changes described in \cref{sec:method-changes}.

\subsection{Summary of Method}
\label{sec:method-summary}

We now explain our model in more detail. We wish to determine the dust extinction along a sightline, using broadband photometry from stars that lie in a small angular region of the sky. In order to infer the extinction along the sightline, we must model the most important factors that go into producing the stellar photometry we observe: the type of each star, the distance to each star, the distribution of dust along the sightline, the variation in dust column density with angle across the sightline, and the photometric uncertainties in the stellar photometry. Our variables are:

\begin{itemize}
    \item In each angular pixel $j$, the dust density in each distance bin $k$: $\alpha_{jk}$, $k = 1, \ldots , n_{\mathrm{distances}}$. Dust column density is related to increase in extinction in each band by an assumed universal extinction vector, $\vect{R}$.
    \item For each star $i$ in angular pixel $j$, the stellar type, $\vect{\Theta}_{ji}$, distance modulus, $\mu_{ji}$, and fractional offset, $\delta_{ji}$, from the ``average'' reddening at an equivalent distance in the pixel.
\end{itemize}

Our model of how our data is generated is as follows:

\begin{enumerate}
    \item Stellar angular positions are fixed. Split the sky up into small sightlines (labeled by $j$) and split each sightline into a fixed number of logarithmically spaced distance bins (labeled by $k$). The sightlines will be treated independently, and the following steps apply to each individual sightline $j$.
    \item For each star $i$, draw intrinsic stellar parameters $\Theta_{ji}$ (luminosity, metallicity) and distance modulus $\mu_{ji}$ from a smooth prior on the distribution of stars of different types throughout the Galaxy.
    \item Draw the line-of-sight dust density $\alpha_{jk}$ (i.e., dust density in each distance bin $k$) from a smooth prior on distribution of dust throughout the Galaxy. This determines the relation between distance modulus and cumulative extinction in the sightline, $\vect{A}_{j} \arg{\mu}$.
    \item Calculate the extinction, $\vect{A}_{ji}$, of each star $i$ in sightline $j$. This is obtained from the line-of-sight extinction, $\vect{A}_{j} \arg{\mu_{ji}}$, with a fractional offset, $\delta_{ji}$, particular to star $i$.
    \item The absolute magnitude, $\vect{M}_{ji}$ (a vector containing one entry per passband), of star $i$ is a deterministic function of stellar type, $\Theta_{ji}$.
    \item The apparent magnitude, $\vect{m}_{ji}$ (also a vector), of star $i$ is the sum of its absolute magnitude, $\vect{M}_{ji}$, distance modulus, $\mu_{ji}$, and extinction, $\vect{A}_{ji}$, with the addition of Gaussian observational noise. The apparent magnitudes of the stars are our observed quantity.
\end{enumerate}

We thus infer stellar type $\Theta_{ji}$, stellar distance modulus $\mu_{ji}$ and line-of-sight dust density $\alpha_{jk}$ from the observed apparent magnitudes $\vect{m}_{ji}$, where $j$ indexes the sightlines, $i$ indexes the stars within each sightline, and $k$ indexes the distance bins within each sightline. This model is shown as a directed factor graph (a type of plate diagram, and described in \citealt{Dietz2010}) in \cref{fig:bayestar-plate-diagram}.

In terms of probability densities, our model can be described as follows. The extinction of star $i$ (in sightline $j$) is determined by $\mu_{ji}$, $\delta_{ji}$ and $\left\{ \alpha_{jk} \right\}_{\forall k}$ (i.e., dust density in all distance bins). Together with the distance modulus $\mu_{ji}$ and type $\Theta_{ji}$ of the star, one can obtain a model apparent magnitude (in each passband), $\vect{m}_{ji}$, for the star, and thus a likelihood:
\begin{align}
    p \arg{\vect{m}_{ji} \, | \, \mu_{ji} , \Theta_{ji} , \delta_{ji} , \left\{ \alpha_{jk} \right\}_{\forall k}} \, .
\end{align}
We also have per-star priors on distance and stellar type, given by a smooth model of the distribution of stars throughout the Galaxy, and a prior on the offset of each star from the local reddening column:
\begin{align}
    p \arg{\mu_{ji} , \Theta_{ji} , \delta_{ji} \, | \, \left\{ \alpha_{jk} \right\}_{\forall k}}
    &= p \arg{\mu_{ji} , \Theta_{ji}}
    \notag \\ & \hspace{0.5cm} \times
    p \arg{\delta_{ji} \, | \, \mu_{ji} , \left\{ \alpha_{jk} \right\}_{\forall k}} \, .
\end{align}
Finally, we have a log-normal prior on the dust density in each distance bin $k$ along each sightline $j$, $p \arg{\left\{ \alpha_{jk} \right\}_{\forall k}}$, whose mean is chosen to match a smooth model of the distribution of dust throughout the Galaxy.

The posterior on the line-of-sight reddening along one sightline $j$ is then given by
\begin{align}
    p \arg{\left\{ \alpha_{jk} \right\}_{\forall k} \, | \left\{ \vect{m}_{ji} \right\}_{\forall i}}
    &= p \arg{\left\{ \alpha_{jk} \right\}_{\forall k}}
    \notag \\ & \hspace{0.25cm} \times
    \prod_{i=1}^{n_{\mathrm{stars}}}
    p \arg{\vect{m}_{ji} \, | \, \mu_{ji} , \Theta_{ji} , \delta_{ji} , \left\{ \alpha_{jk} \right\}_{\forall k}}
    \notag \\ & \hspace{0.25cm} \times
    p \arg{\mu_{ji} , \Theta_{ji}} p \arg{\delta_{ji} \, | \, \mu_{ji} , \left\{ \alpha_{jk} \right\}_{\forall k}} \, .
\end{align}
We pre-compute the likelihood and prior terms for the individual stars, and then sample in $\alpha_{jk}$. The mathematical details of the model and of our sampling method are given in \citet{Green2014-method,Green2015-release}.

\subsection{Changes to Bayestar15 Approach}
\label{sec:method-changes}

\begin{table}
    \caption{Extinction vector adopted in this work.}
    \label{tab:extinction-vector}
    \centering
    \begin{footnotesize}
        \setlength{\tabcolsep}{0.5em}
        \begin{tabular}{c c c c c c c c}
            \hline
            \hline \\[-10pt]
             $\gps$ & $\rps$ & $\ips$ & $\zps$ & $\yps$ & $J$ & $H$ & $K_{\mathrm{S}}$ \\ \\[-10pt]
            \hline \\[-10pt]
            3.384  &  2.483  &  1.838  &  1.414  &  1.126  &  0.650  &  0.327  &  0.161  \\[2pt]
            \hline
        \end{tabular}
    \end{footnotesize}
\end{table}

While the method used in this paper is substantially the same as that used in Bayestar15, we have introduced a number of changes.

First, we use a new extinction vector, relating dust density to the extinction in each passband. In Bayestar15, we used an extinction vector based on \citet{Fitzpatrick1999} (calculated for PS1 passbands by \citealt{Schlafly2011}) and \citet{Cardelli1989} (calculated for 2MASSS passbands by \citealt{Yuan2013}). In this work, we make use of an extinction vector derived by comparison of spectra from the APO Galactic Evolution Experiment \citep[``APOGEE'']{APOGEE} and the broadband colors of stars \citep[``S16'']{Schlafly2016-extcurve}. As a significant fraction of APOGEE targets lie at low Galactic latitudes, the extinction vector found by S16 is more appropriate to the Galactic plane -- where most of the dust lies -- than that found by \citet{Schlafly2011}. In addition, S16 derives the extinction vector across optical and NIR passbands simultaneously, leading to a more consistent treatment of extinction across our full wavelength range. By modeling the colors of stars as a function of spectral type and a reddening component, S16 derives a mean dust extinction vector, and determines the principal axes in color space along which the extinction vector varies. This is a variant of the ``pair method,'' where the dust extinction vector is determined by comparing stars of the same spectral type (and therefore, the same luminosity and color) that lie behind different amounts of dust obscuration (\citealt{Trumpler1930-pair-method}; for more recent examples of this technique, see \citealt{Schlafly2011} and \citealt{Yuan2013}). The extinction vector derived by S16 has an undetermined gray component -- that is, an arbitrary scalar constant can be added into the extinction vector. In addition, the overall scaling of the extinction vector (i.e., the relation between extinction and dust density) is undetermined. In mathematical language, S16 determines a vector, $\vect{R}_0$, which for a given dust column density, $\sigma$, is related to extinction, $\vect{A}$, by
\begin{align}
    \vect{A} = \sigma a \left( \vect{R}_0 + b \right) \, ,
\end{align}
where $a$ and $b$ are unknown scalar constants. We set the gray component of the reddening vector, $b$, such that the extinction in the WISE $W2$ passband is zero, regardless of dust column density. This should bring us close to the correct value of $b$, as dust optical depth in the $W2$ passband is small compared to that in optical passbands.

Our method depends trivially on the value of $a$, since any scaling in the overall extinction vector can be compensated by a reciprocal scaling in the dust column density. The parameter $\alpha_{jk}$, which represents the amount of dust in distance bin $k$ of sightline $j$ (see \cref{fig:bayestar-plate-diagram}), gives the quantity of dust reddening in an arbitrary unit. By multiplying these units by the extinction vector, $\vect{R}$, we arrive at a physically meaningful unit -- extinction in each passband.

\cref{tab:extinction-vector} gives the components of our adopted extinction vector, which is based on S16. In order to transform values in our Bayestar17 map into extinction in any PS1 or 2MASS passband, one should multiply our map value by the appropriate component of the extinction vector.

If instead of assuming that $A_{W2} = 0$, we use the result from \citet{Indebetouw2005} that $A_{H} / A_{K} = 1.55$ to set the constant $b$, we arrive at $A_{W2} = 0.12 \, \mathrm{mag}$ for $\Erz = 1 \, \mathrm{mag}$. This implies a possible bias in distance of 5\% at $\Erz = 1 \, \mathrm{mag}$ (similar to $\EBV = 1 \, \mathrm{mag}$), and 10\% at $\Erz = 2 \, \mathrm{mag}$.

We also use newer PS1 data (see \cref{sec:ps1}), which requires the creation of new stellar templates, which give the absolute magnitude of a star in the PS1 and 2MASS passbands as a function of luminosity and metallicity, suited to the latest calibration of the PS1 point-source catalog. These templates were generated according to the same technique described in \citet{Green2014-method,Green2015-release}.

The model described in \cref{fig:bayestar-plate-diagram} does not specify \textit{how} the sky is pixelized into lines of sight. There, the assignment of stars to sightlines is taken as a given. In Bayestar15 and in this work, we use the HEALPix pixelization scheme \citep{Gorski2005}. In Bayestar15, pixel size varied over the sky, and was determined by a heuristic that took into account observed density of stars. Beginning with large HEALPix pixels ($\mathtt{nside} = 64$, corresponding to an angular scale of $\sim \! 55^{\prime}$), we recursively subdivided each pixel until the number of stars fell below a given threshold (given here in \cref{tab:thresholds}). This threshold was based on the pixel scale, with the number of stars required to trigger a further subdivision increasing as the pixel scale became smaller. The idea behind this scheme is to provide greater angular resolution in areas where there are more stars available to trace the dust column. Across much of the sky, this heuristic works well, allowing us to achieve similar signal-to-noise in regions of the sky with vastly different stellar surface densities. One undesirable outcome of this heuristic, however, is that areas of the sky with particularly dense dust clouds, which block out a significant fraction of the background stars, are pixelized more roughly. Yet these dense clouds contain intricate dust substructure, which our model (uniform extinction across a sightline, with only white-noise angular substructure) captures poorly. Therefore, in the current work, we modify the pixelization procedure of Bayestar15 to make the subdivision threshold depend not only on the pixel scale, but also on the SFD estimate of dust reddening.

In detail, we multiply the subdivision thresholds by $V$-band transparency. If the threshold number of stars for HEALPix nside $m$ is $N_{m}$, then the modified threshold is given by
\begin{align}
    N_{m}^{\prime} = 10^{-\frac{2}{5} A_{V}} \, N_{m} \, ,
\end{align}
where $A_{V} = 3.1 \, \EBV_{\mathrm{SFD}}$ is the $V$-band extinction given by SFD. Here, we use the mean $\EBV_{\mathrm{SFD}}$ of the stars in the pixel. This will have the effect of causing greater subdivision in regions with high optical extinction. We limit the suppression of the subdivision threshold so that $N_{m}^{\prime} \geq \tfrac{1}{4} N_{m}$. As a consequence, the optical transparency term can only cause one additional subdivision. As we show in \cref{sec:comparison-with-bayestar}, our new dust map has finer angular resolution than Bayestar15 in regions of high extinction.

The final two changes we have made to our method are minor. In our prior on the spatial distribution of stars, we have adjusted the local abundance of halo stars, relative to the local abundance of thin-disk stars. In Bayestar15, we had set the local halo fraction to $0.0006$, as indicated by \citet{Robin2003}. However, we found that our inferred distribution of stellar distances indicated a greater local halo fraction. We have therefore increased the local halo fraction to $0.003$, closer to the value suggested by \citet{Juric2008}. We use the same smooth prior on the distribution of dust as in Bayestar15, with one small modification. As before, our prior on the density of dust in each distance bin of each sightline is log-normal, with a mean value set to trace a smooth disk model, described in detail in \S 2.1 of \citet{Green2015-release}. We found previously that setting a lower limit on the mean of the log-normal distribution in each distance bin improved convergence. We have lowered this floor from $\Delta \EBV = e^{-12} \, \mathrm{mag}$ to $\Delta \EBV = e^{-24} \, \mathrm{mag}$ per distance bin. We have found that this change does not significantly affect convergence, and that it allows us to more rigidly enforce the prior that there is little dust far away from the midplane of the Galaxy.

\begin{figure*}
	\includegraphics[width=\textwidth]{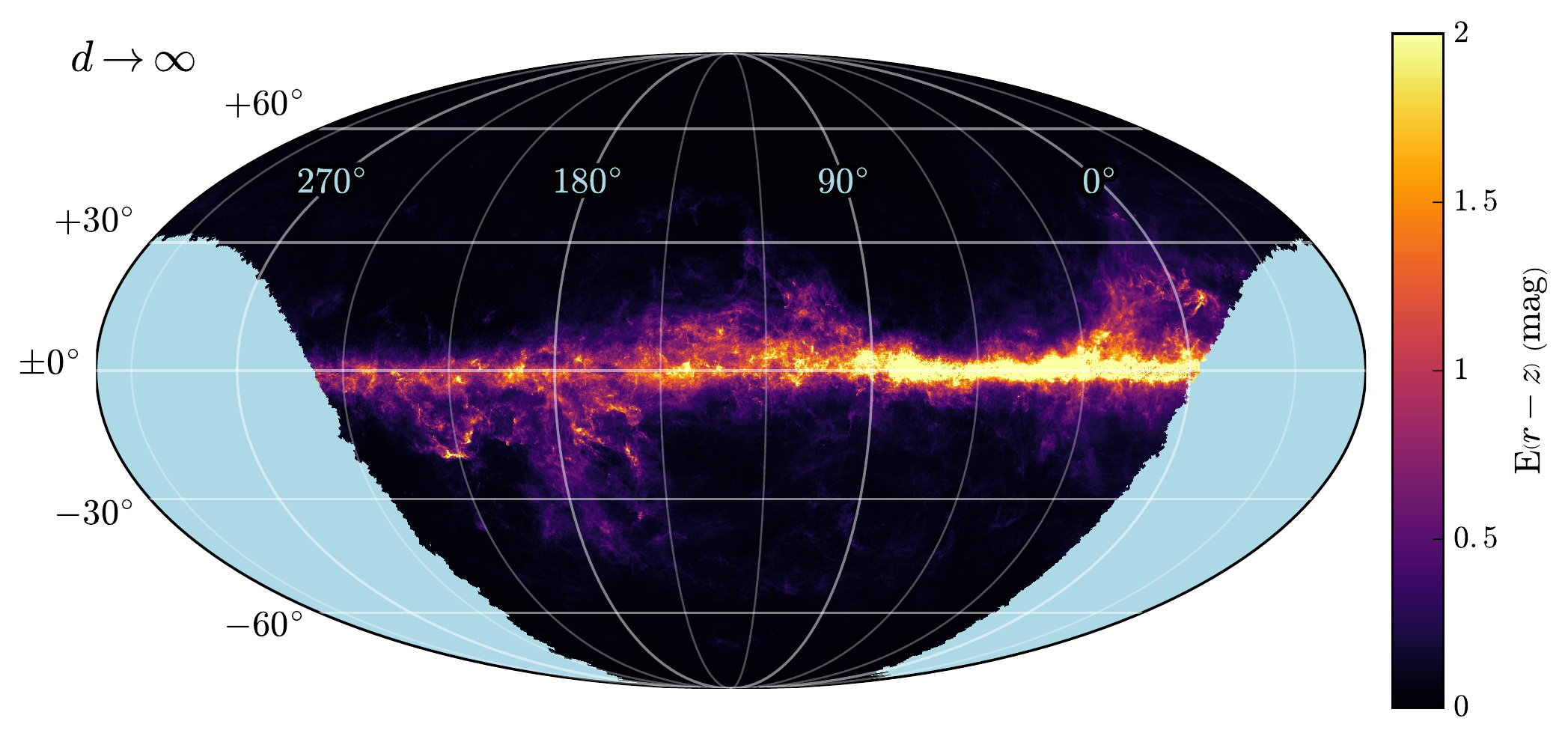}
    \caption{Cumulative reddening out to the maximum distance in the 3D
             dust map, in magnitudes of \Erz. The map uses a Galactic
             Mollweide projection, with the $\ell = \deg{0}$ meridian
             rotated to the right of the map. The map covers declinations
             north of $-\deg{30}$ (the PS1 $3\pi$ survey footprint).
             Regions shaded blue lie south of declination $-\deg{30}$,
             and are therefore not covered by this map.}
    \label{fig:bmk_cumulative}
\end{figure*}

\begin{figure*}
	\includegraphics[width=\textwidth]{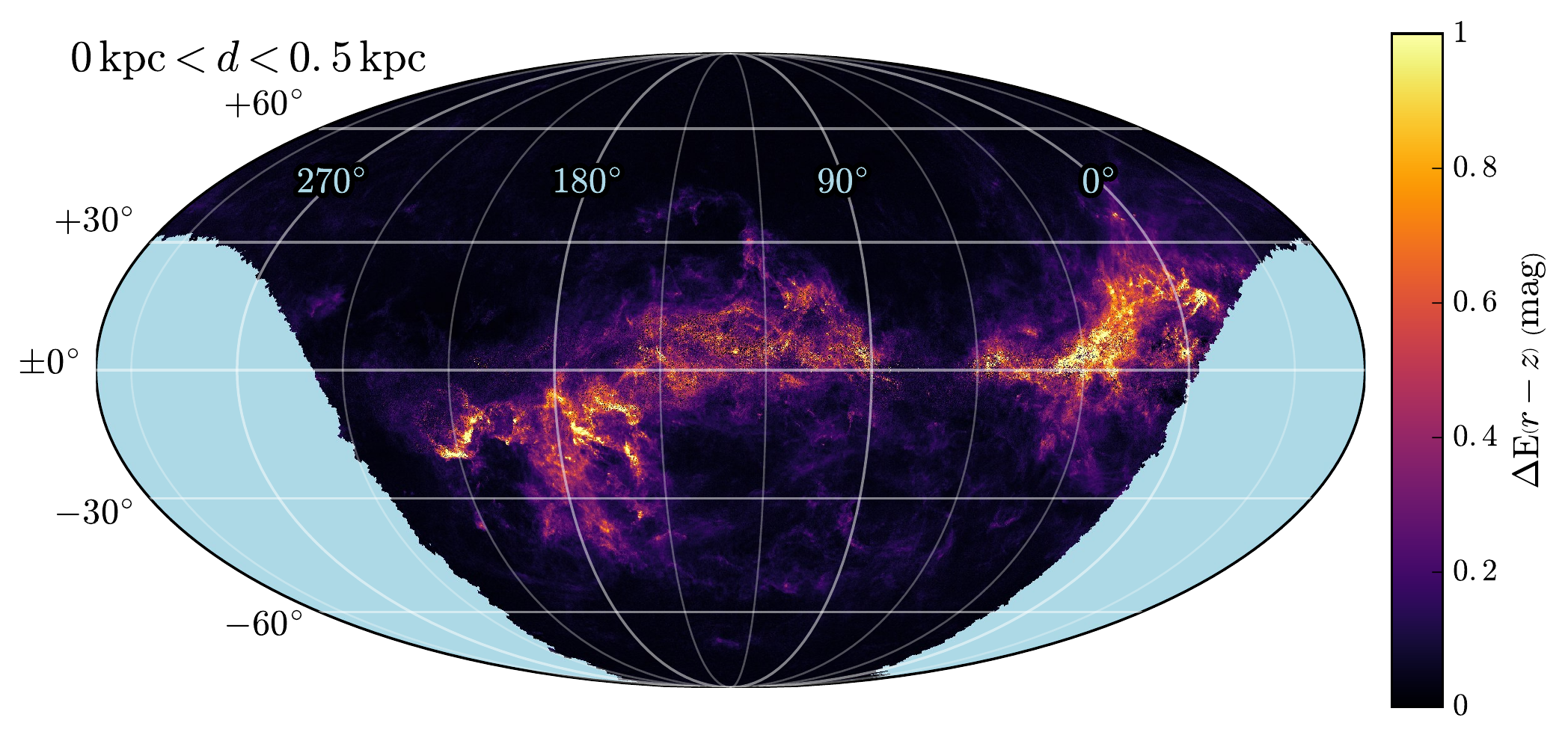}
    \caption{Cumulative reddening out to 500~pc, in magnitudes of \Erz.}
    \label{fig:bmk_diff_00}
\end{figure*}

\begin{figure*}
	\includegraphics[width=\textwidth]{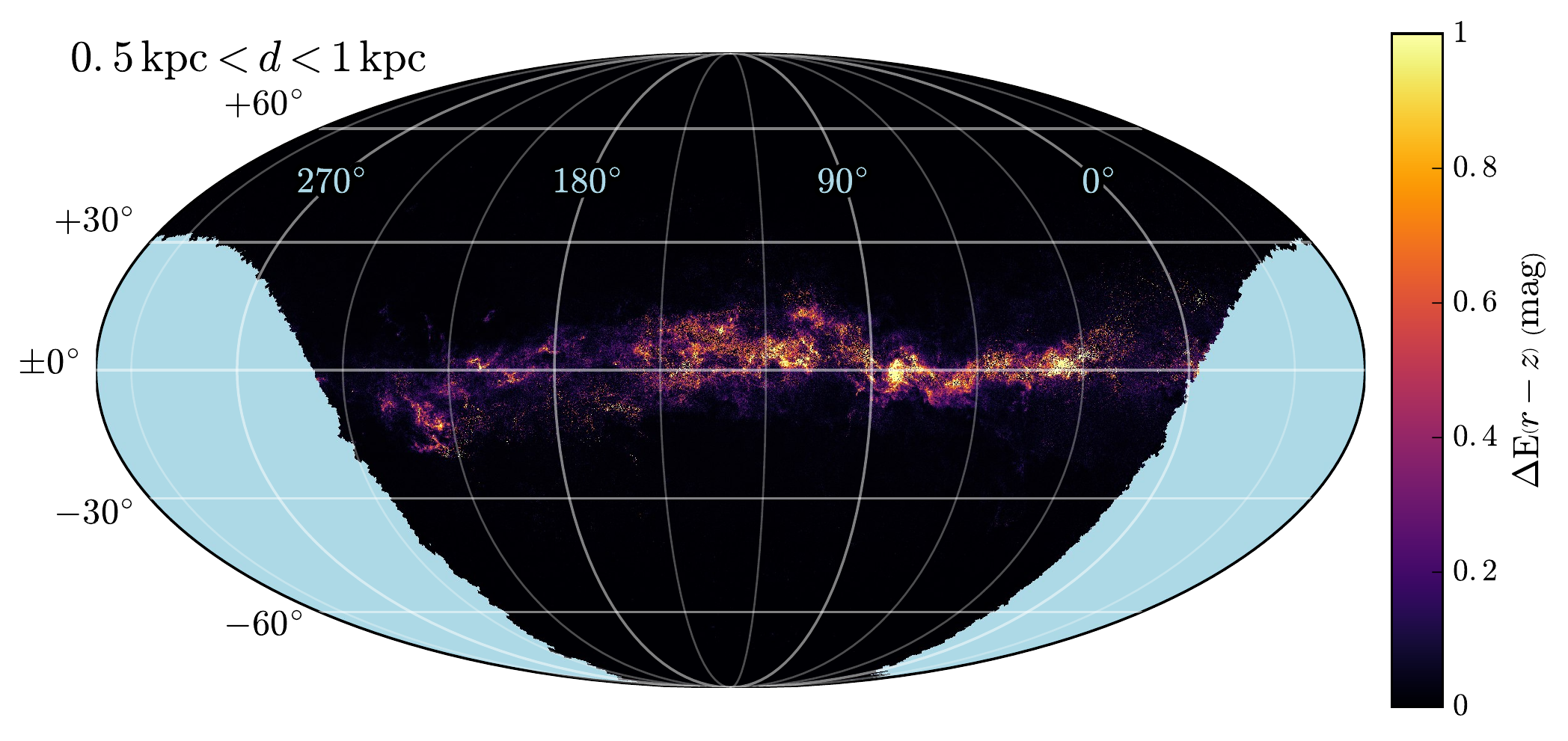}
    \caption{Cumulative reddening between 500~pc and 1~kpc, in magnitudes
             of \Erz.}
    \label{fig:bmk_diff_01}
\end{figure*}

\begin{figure*}
	\includegraphics[width=\textwidth]{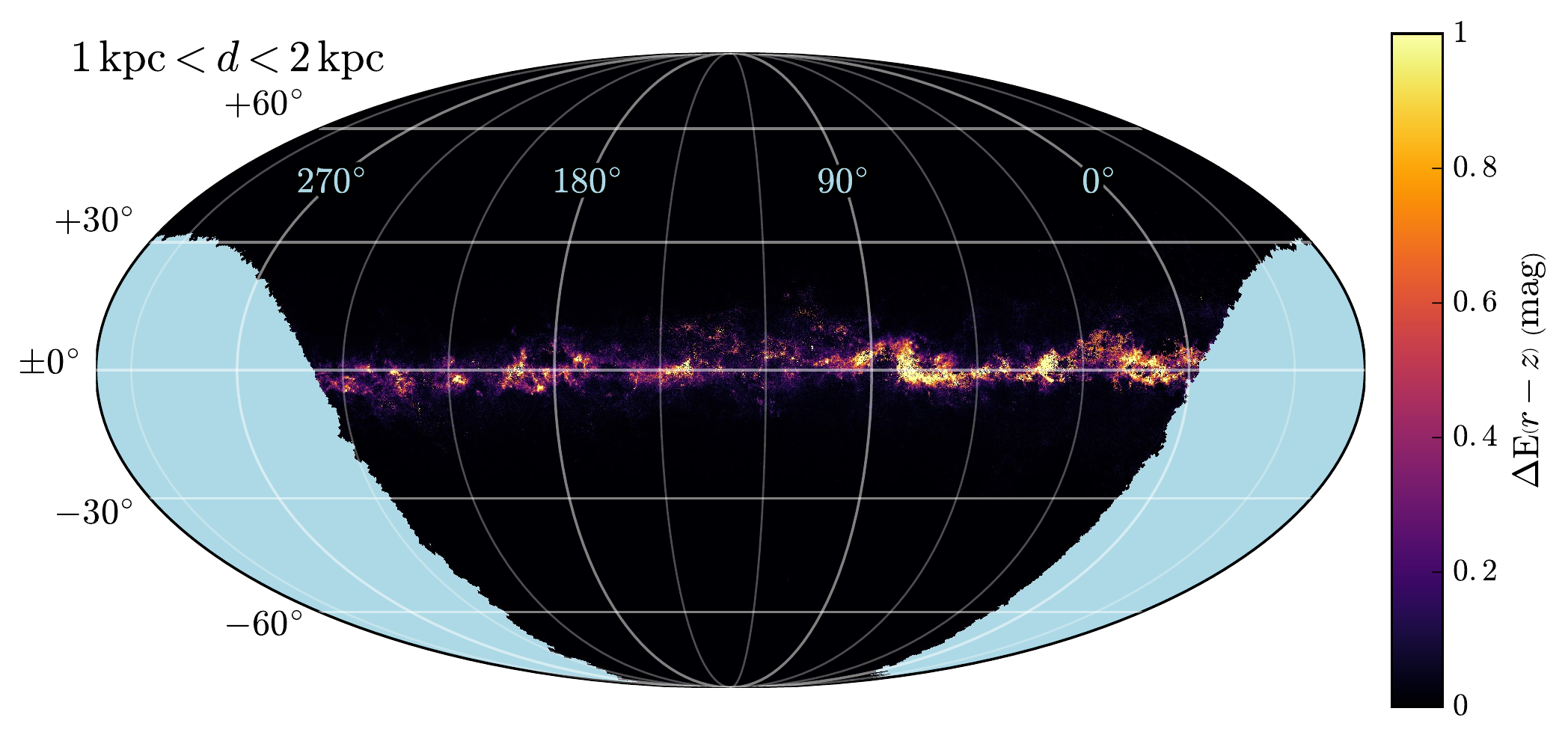}
    \caption{Cumulative reddening between 1~kpc and 2~kpc, in magnitudes
             of \Erz.}
    \label{fig:bmk_diff_02}
\end{figure*}

\begin{figure*}
	\includegraphics[width=\textwidth]{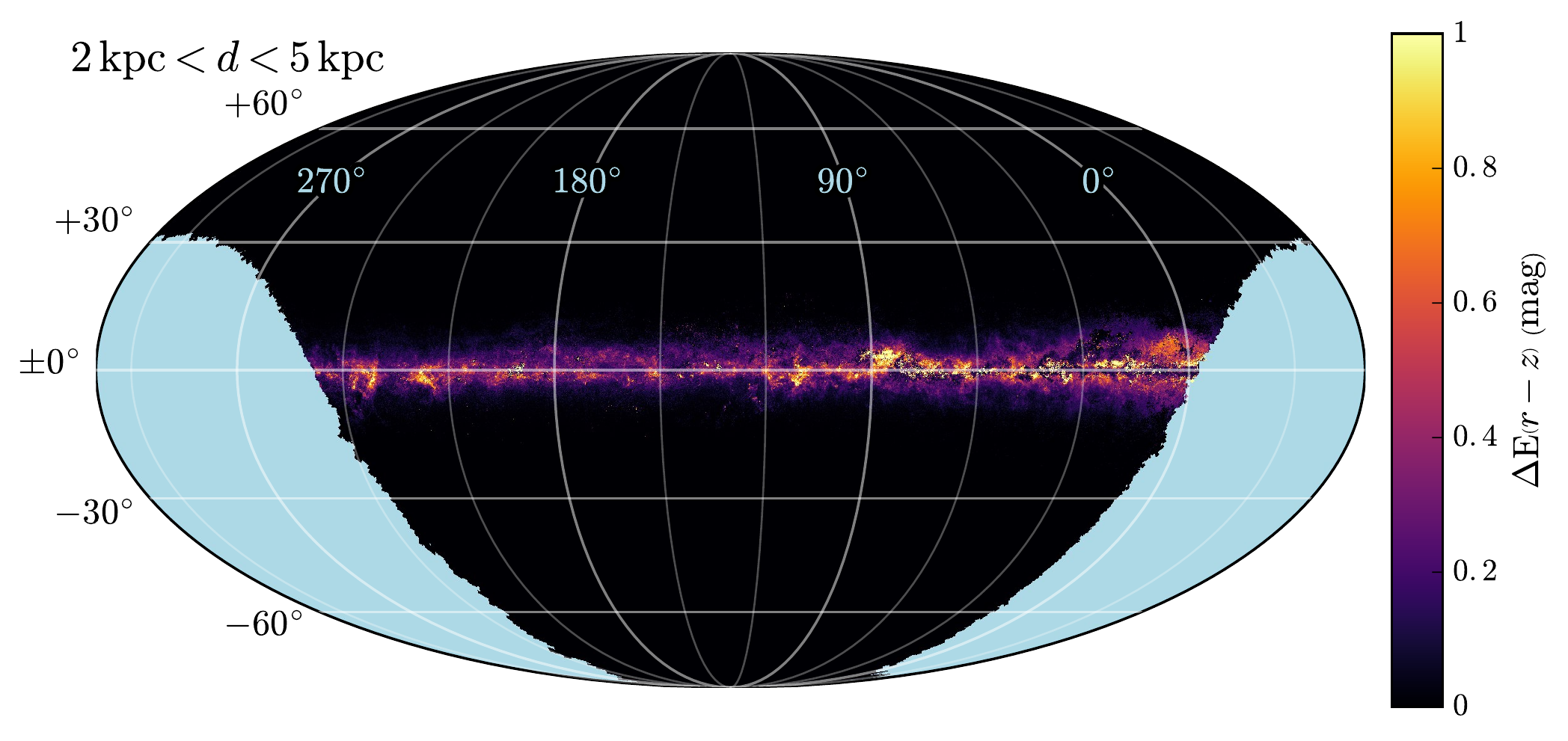}
    \caption{Cumulative reddening between 2~kpc and 5~kpc, in magnitudes
             of \Erz.}
    \label{fig:bmk_diff_03}
\end{figure*}

\begin{figure*}
	\includegraphics[width=0.95\textwidth]{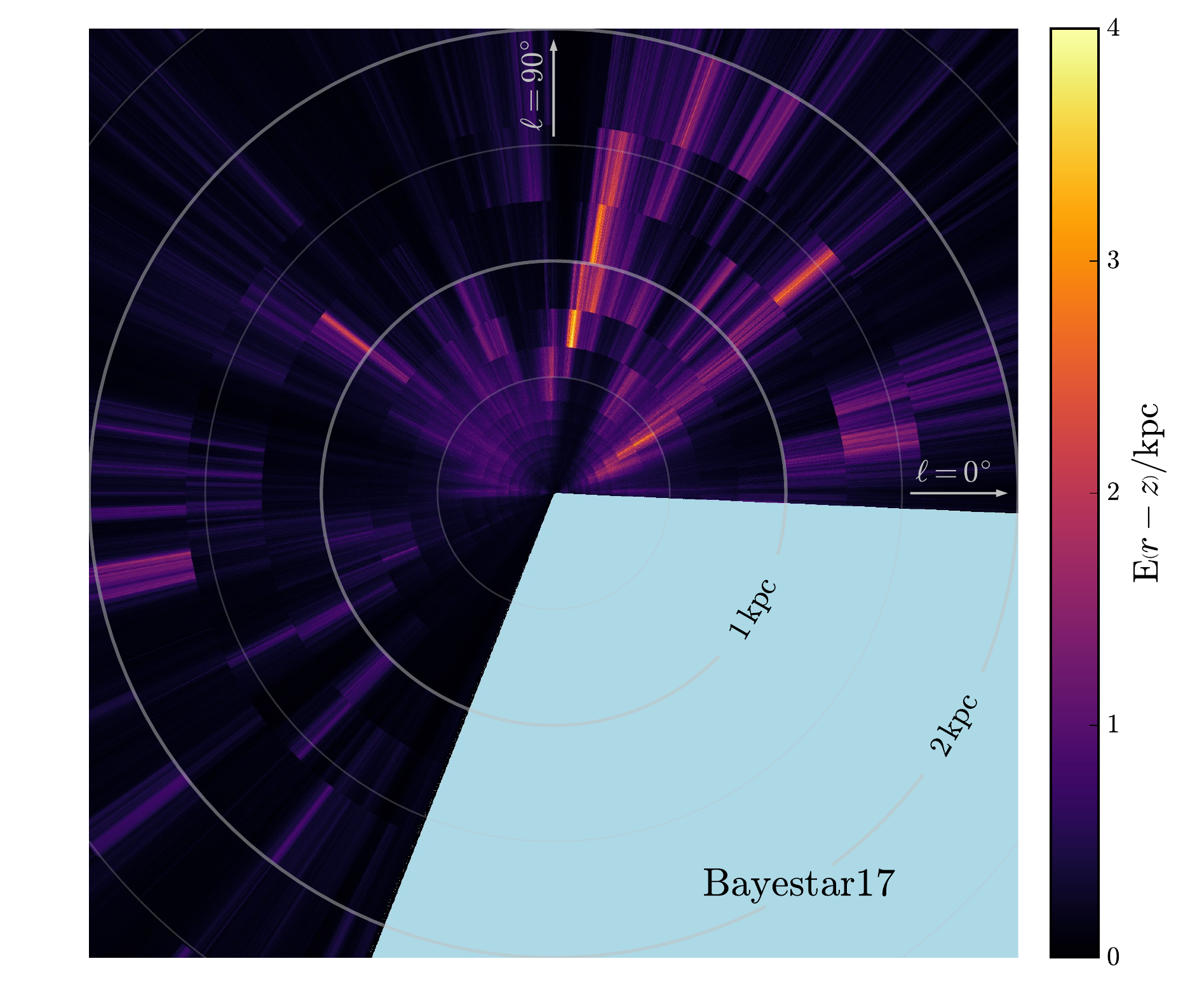}
    \caption{A bird's eye view of our new dust map, showing reddening per kiloparsec in the plane of the Galaxy. At each location in the plane of the Galaxy, mean reddening per kiloparsec within 50~pc of the $b = \deg{0}$ plane is plotted. The Sun is located at the center of plot, and the Galactic center lies off the plot to the right. The region without data lies below declination $\delta = -\deg{30}$. The ``fingers-of-God'' effect, by which clouds are elongated along the radial axis (from the perspective of the Sun), is due to the fact that angles to dust clouds are much better determined than distances.}
    \label{fig:bmk_birdseye}
\end{figure*}

\section{Data}
\label{sec:data}

As in Bayestar15, we use broadband photometry from two large photometric surveys, Pan-STARRS~1 (PS1) and the Two Micron All-Sky Survey (2MASS).

\subsection{Pan-STARRS 1}
\label{sec:ps1}

The Panoramic Survey Telescope and Rapid Response System 1 (Pan-STARRS~1, or PS1) is a 1.8-meter optical and near-infrared telescope located on Mount Haleakala, Hawaii \citep{Chambers2016-PS1}. The telescope is equipped with the Gigapixel Camera \#1 (GPC1), consisting of an array of 60 CCD detectors, each 4800 pixels on a side \citep{Tonry2006-GPC1,Onaka2008-GPC1,Chambers2016-PS1}. From May 2010 to April 2014, the majority of the observing time was dedicated to a multi-epoch $3 \pi$-steradian survey of the sky north of declination $\delta = -\deg{30}$ \citep[the ``$3 \pi$ survey'']{Chambers2016-PS1}. The $3 \pi$ survey observed in five passbands, $\gps$, $\rps$, $\ips$, $\zps$ and $\yps$, similar to those used by SDSS \citep{York2000}, with the most significant difference being the replacement of the Sloan $u$ band with a near-infrared band, $\yps$. The PS1 filter set spans 400--1000~nm \citep{Stubbs2010-PS1-lasercal}. The images are processed by the PS1 Image Processing Pipeline (IPP) \citep{Magnier2016-PS1-pixel-analysis}, which performs automatic astrometry and photometry \citep{Magnier2016a-PS1-photo-astro-calib}. The internal relative photometric calibration of the data is better than 1\% \citep{Chambers2016-PS1,Schlafly2012}. The $3 \pi$ survey reaches typical single-epoch $5 \sigma$ depths of 22.0~mag (AB) in $\gps$, 21.8~mag in $\rps$, 21.5~mag in $\ips$, 20.9~mag in $\zps$, and 19.7~mag in $\yps$ \citep{Chambers2016-PS1}. The resulting homogeneous optical and near-infrared coverage of three quarters of the sky makes the PS1 photometry ideal for studies of the distribution of the Galaxy's dust.

The PS1 Data Release 1 (DR1) occurred in December 2016. Among other data products, DR1 included the static sky catalog and stack images from the $3 \pi$ survey \citep{Flewelling2016-PS1-database}.

As compared to Bayestar15, we use an additional $\sim$1.5 years of data from PS1. This data includes the north equatorial pole, which was not included in the data reduction used by Bayestar15. The present paper is based on single-epoch data from the PS1 DR1 $3 \pi$ survey.

\subsection{Two Micron All Sky Survey}
\label{sec:2mass}

The Two Micron All Sky Survey (2MASS) is a uniform all-sky survey in three near-infrared bandpasses, $J$, $H$ and $K_{s}$ \citep{Skrutskie2006}. The survey derives its name from the wavelength range covered by the longest-wavelength band, $K_{s}$, which lies in the longest-wavelength atmospheric window not severely affected by background thermal emission \citep{Skrutskie2006}. The survey was conducted from two 1.3-meter telescopes, located at Mount Hopkins, Arizona and Cerro Tololo, Chile, in order to provide coverage for both the northern and southern skies, respectively. The focal plane of each telescope was equipped with three $256 \times 256$ pixel arrays, with a pixel scale of $2^{\prime \prime} \times 2^{\prime \prime}$. Each field on the sky was covered six times, with dual 51-millisecond and 1.3-second exposures, achieving a $10 \sigma$ point-source depth of approximately 15.8, 15.1 and 14.3~mag (Vega) in $J$, $H$ and $K_{s}$, respectively. Calibration of the survey is considered accurate at the 0.02~mag level, with per-source photometric uncertainties for bright sources below 0.03~mag \citep{Skrutskie2006}.

We use the 2MASS ``high-reliability catalog,'' requiring in addition that there be no contamination/confusion with neighboring point sources or with galaxies.

\subsection{Input Catalog}
\label{sec:input-catalog}

\begin{table}
    \caption{Pixelization of the sky.}
    \label{tab:thresholds}
    \centering
    \begin{footnotesize}
        \setlength{\tabcolsep}{0.75em}
        \begin{tabular}{c c c c r}
            \hline
            \hline \\[-10pt]
            \multirow{2}{*}{\texttt{nside}} &
            pixel &
            Max. &
            $\Omega$ &
            \# of \\
            &
            scale &
            Stars/pix &
            ($\mathrm{deg}^{2}$) &
            Pixels \\ \\[-10pt]
            \hline \\[-10pt]
            64    &  $55^{\prime}$  &  200  &  62     &  74       \\
    		128   &  $27^{\prime}$  &  250  &  86     &  410      \\
    		256   &  $14^{\prime}$  &  300  &  10080  &  192156   \\
    		512   &  $6.9^{\prime}$  &  800  &  14179  &  1080495  \\
    		1024  &  $3.4^{\prime}$  &  ---  &  7041   &  2147770  \\[2pt]
            \hline \\[-10pt]
    		total &  ---  &  ---  &  31439  &  3420905  \\[2pt]
            \hline
        \end{tabular}
    \end{footnotesize}
\end{table}

\begin{table}
    \caption{Source statistics.}
    \label{tab:n-passbands}
    \centering
    \begin{footnotesize}
        \setlength{\tabcolsep}{0.5em}
        \begin{tabular}{c c c c c}
            \hline
            \hline \\[-10pt]
            \multicolumn{2}{c}{\% sources detected} & & \multicolumn{2}{c}{\% sources detected} \\
            \multicolumn{2}{c}{in each band} & & \multicolumn{2}{c}{in $N$ bands} \\
            \hline \\[-10pt]
             $\gps$           & 78.9\% & & 4  & 28.9\% \\
             $\rps$           & 98.1\% & & 5  & 52.5\% \\
             $\ips$           & 97.0\% & & 6  & 4.8\%  \\
             $\zps$           & 98.1\% & & 7  & 5.1\%  \\
             $\yps$           & 93.9\% & & 8  & 8.7\%  \\
             $J$              & 18.7\% & & -- & --     \\
             $H$              & 15.6\% & & -- & --     \\
             $K_{\mathrm{S}}$ & 11.8   & & -- & --     \\ \\[-10pt]
            \hline \\[-10pt]
            \multicolumn{5}{l}{806 million sources total.} \\
            \hline
        \end{tabular}
    \end{footnotesize}
\end{table}

We match 2MASS to PS1 sources, requiring detection in at least two PS1 passbands, and at least four passbands in total. Our input catalog contains 806 million sources, divided among 3.42 million sightlines. Over 80\% of sources are detected in four or five passbands, with the remaining 20\% of sources having detections in 6, 7 or 8 passbands. After adaptive subdivision of the sky, the number of pixels at each angular scale (or, equivalently, HEALPix \texttt{nside}) is given in \cref{tab:thresholds}.

\section{Results}
\label{sec:results}

We massively parallelized our computations across the Harvard Odyssey compute cluster, taking advantage of the independence of each sightline in our model. To check for convergence, we ran four independent MCMC chains for each sightline. The computations took a total of 2.5 million CPU hours, amounting to $2.7 \, \mathrm{s} / \mathrm{star} / \mathrm{chain}$. The final product is a 3D map of interstellar reddening, covering three quarters of the sky at typical resolutions ranging from $3.4^{\prime}$ to $13.7^{\prime}$.

In \cref{sec:all-sky-maps}, we present maps of both cumulative reddening and differential reddening in discrete distance slices. In \cref{sec:comparison-with-bayestar}, we compare our new 3D dust map with our previous map \citep{Green2015-release}. In \cref{sec:planck-comparison}, we compare our new 3D dust map with 2D dust maps derived from far-infrared dust emission.

\subsection{$3 \pi$ Steradian Maps of Reddening}
\label{sec:all-sky-maps}

\cref{fig:bmk_cumulative} shows cumulative reddening across the PS1 footprint, covering the $3\pi$ steradians of the sky that lies north of declination $\delta = -\deg{30}$. As can be seen, we recover the familiar projected dust structure. \cref{fig:bmk_diff_00,fig:bmk_diff_01,fig:bmk_diff_02,fig:bmk_diff_02,fig:bmk_diff_03} show the differential reddening in discrete distance slices. As we reside in the plane of a disk galaxy, dust extends to the greatest latitudes in the nearest distance slice. In the nearest distance panel, \cref{fig:bmk_diff_00}, well-known nearby clouds are visible, such as the Perseus-Taurus-Auriga cloud complex ($\deg{150} \lesssim \ell \lesssim \deg{190}$, $-\deg{40} \lesssim b \lesssim -\deg{10}$), Cepheus ($\ell \approx \deg{110}$, $b \approx \deg{17}$), the $\rho$ Ophiuchi cloud complex ($\ell \approx \deg{350}$, $b \approx \deg{15}$), the Orion nebulae ($\ell \approx \deg{210}$, $b \approx -\deg{18}$), the Aquila Rift ($\deg{0} \lesssim \ell \lesssim \deg{30}$, $\deg{0} \lesssim b \lesssim \deg{20}$) and the $\rho$ Ophiuchi cloud complex ($\ell \approx \deg{350}$, $b \approx \deg{15}$). In subsequent distance slices, distinct features are visible, indicating that we have separated out real dust structures at each distance.

\cref{fig:bmk_birdseye} shows a top-down view of Bayestar17, looking straight down on the Galactic plane, with the Solar system at the center of the image. Specifically, the figure shows the mean reddening per kiloparsec within 50~pc of the $b = \deg{0}$ plane.

\subsection{Comparison with Previous Maps}
\label{sec:comparison-with-previous}

There are two different classes of dust maps we would like to compare our results with: 3D dust maps based on stellar photometry, and 2D dust maps based on the optical depth of dust emission in the far infrared. Here, we pick one map of each type to compare against. First, in \cref{sec:comparison-with-bayestar}, we compare our new 3D dust map with the previous version, Bayestar15. Then, in \cref{sec:iphas-comparison}, we compare our map with the 3D dust map of \citet[``Sale14'']{Sale2014b}, which covers the northern Galactic plane. Finally, in \cref{sec:planck-comparison}, we compare our projected map with the 2D dust FIR optical-depth based map from \citet[``Planck14'']{PlanckCollaboration2013}.

\subsubsection{Units of Comparison}
\label{sec:units-of-comparison}

When comparing reddening or extinction maps that make different assumptions about the wavelength--extinction relation, one has to be careful about defining what units are to be used in the comparison. For example, assume that in some arbitrary color, $X - Y$, the reddening estimates given by maps $A$ and $B$ are related by a simple multiplicative factor:
\begin{align}
    \Eab{X}{Y}_{A} = \gamma \, \Eab{X}{Y}_{B} \, ,
\end{align}
where $\gamma$ is some constant. In a new color, $X^{\prime} - Y^{\prime}$, the relation between the two maps will be
\begin{align}
    \Eab{X^{\prime}}{Y^{\prime}}_{A} = \gamma^{\prime} \, \Eab{X^{\prime}}{Y^{\prime}}_{B} \, ,
\end{align}
where
\begin{align}
    \gamma^{\prime} = \left[ \left.
        \frac{ \Eab{X^{\prime}}{Y^{\prime}}_{A} }{ \Eab{X}{Y}_{A} }
        \middle/
        \frac{ \Eab{X^{\prime}}{Y^{\prime}}_{B} }{ \Eab{X}{Y}_{B} }
        \right. \right] \, .
\end{align}
If maps $A$ and $B$ assume different wavelength--extinction relations, then the slope of the linear relation between the two maps depends on the color chosen for the comparison.

In the rest of this section, we conduct our comparisons with the $\rps - \zps$ color. We choose this color because stellar $\EBV$ is approximately equal to $\Erz$ for $R_{V} \approx 3.1$ extinction relations, and because $\rps$ and $\zps$ span the mid-range of wavelengths covered by PS1. The vast majority of stars we use in constructing our 3D dust map are observed in $\rps$, $\ips$ and $\zps$.

\subsubsection{Comparison with Bayestar15}
\label{sec:comparison-with-bayestar}

\begin{figure*}
	\includegraphics[width=\textwidth]{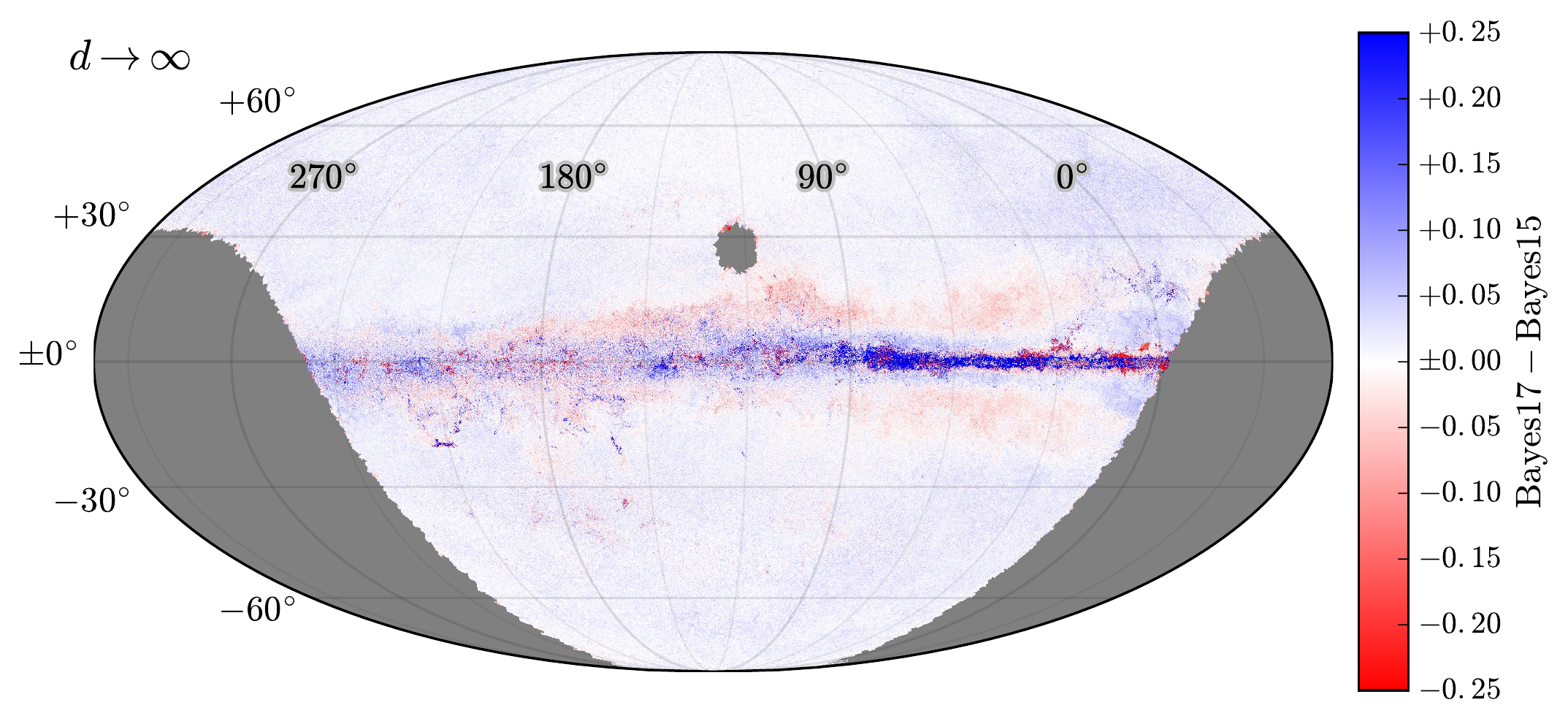}
    \caption{An all-sky map of the residuals between Bayestar17 and Bayestar15, in magnitudes of \Erz. Both maps have been integrated to infinite distance. In each pixel, a random sample from each map is drawn. Positive values (blue) indicate that Bayestar17 infers greater cumulative reddening. Gray corresponds to areas with no PS1 data. The hole at $\ell \approx \deg{120}$, $b \approx \deg{30}$ corresponds to the equatorial north pole, which had not yet been reduced in the version of the PS1 photometry used by Bayestar15.}
    \label{fig:bmk_minus_bayestar_allsky}
\end{figure*}

\begin{figure*}
    \centering
    \begin{subfigure}[t]{0.49\textwidth}
        \centering
    	\includegraphics[width=\linewidth]{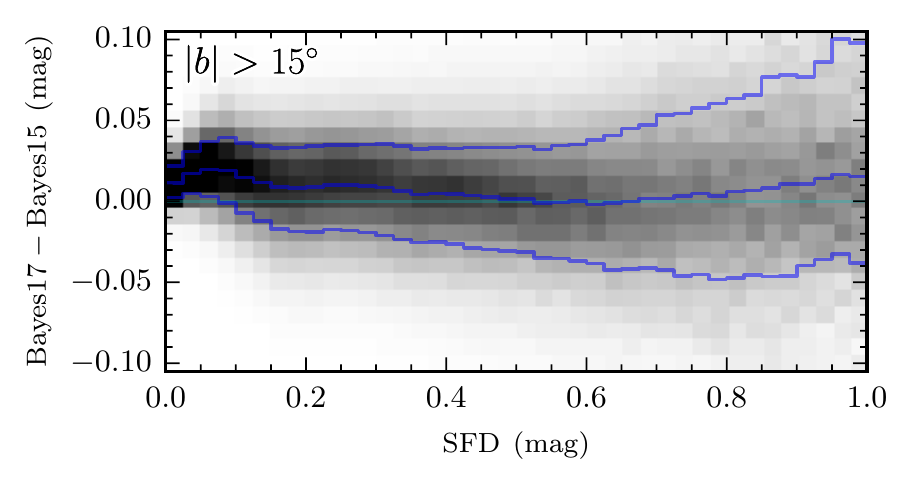}
        \subcaption{}
        \label{fig:corr_sfd_vs_bmk_minus_bayestar}
    \end{subfigure}
    \begin{subfigure}[t]{0.49\textwidth}
        \centering
    	\includegraphics[width=\linewidth]{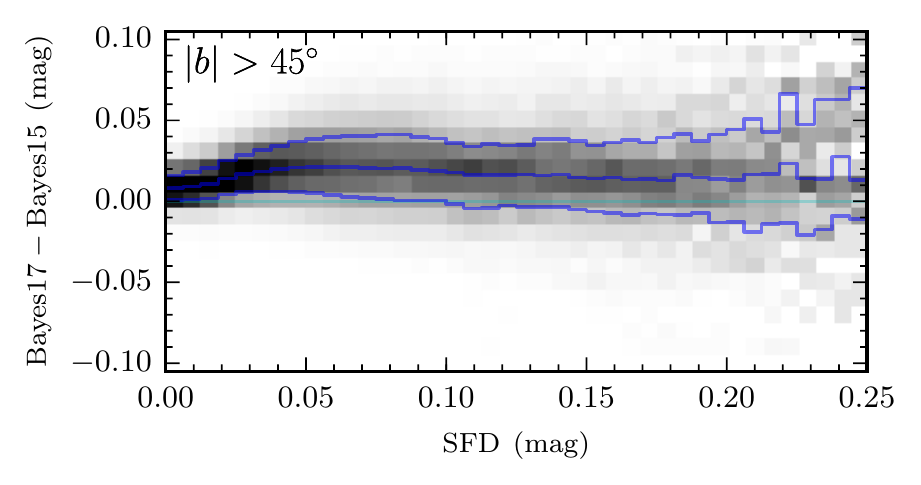}
        \subcaption{}
        \label{fig:corr_sfd_vs_bmk_minus_bayestar_highlat}
    \end{subfigure}
    \par\bigskip
    \begin{subfigure}[b]{0.49\textwidth}
        \centering
    	\includegraphics[width=\linewidth]{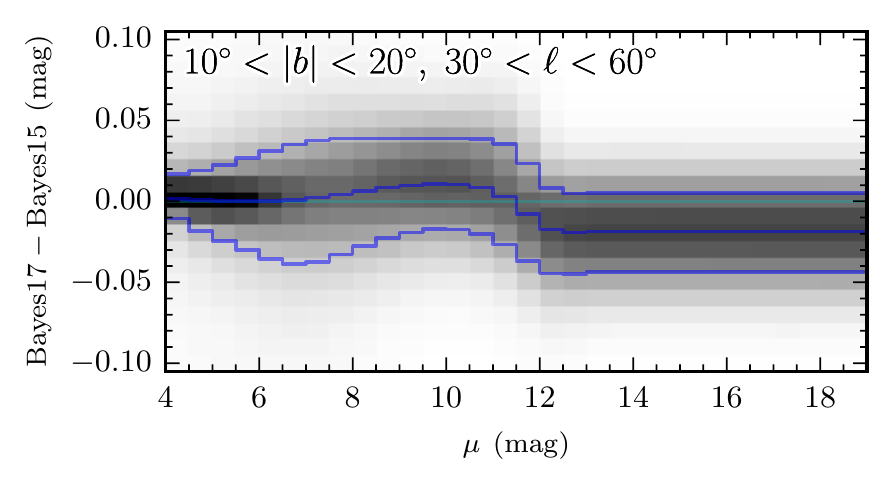}
        \subcaption{}
        \label{fig:corr_distmod_vs_bmk_minus_bayestar_redstrip}
    \end{subfigure}
    \begin{subfigure}[b]{0.49\textwidth}
        \centering
    	\includegraphics[width=\linewidth]{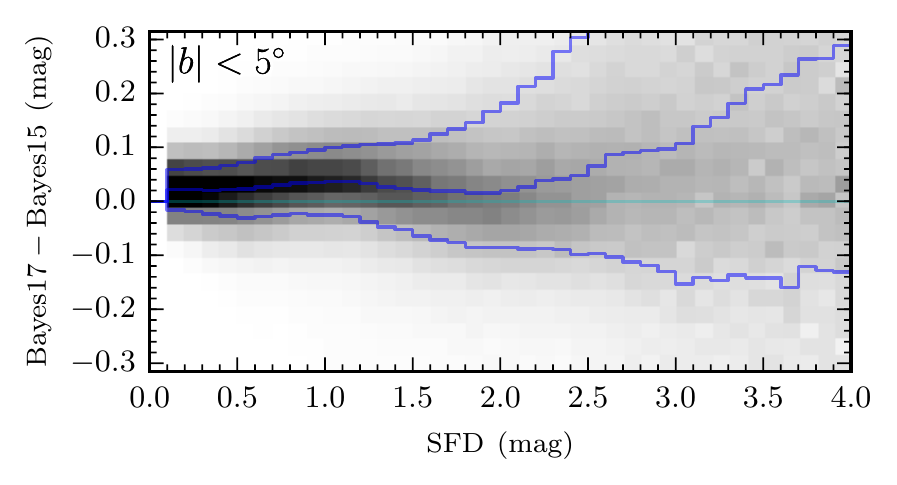}
        \subcaption{}
        \label{fig:corr_sfd_vs_bmk_minus_bayestar_galplane}
    \end{subfigure}

    \caption{Residuals between Bayestar17 and Bayestar15, in magnitudes of \Erz. \textbf{Top left:} As a function of SFD \Erz reddening, off the plane of the Galaxy. For each bin of SFD reddening, we show the histogram of reddening residuals vertically. For both 3D dust maps, we choose random samples of cumulative reddening to compare. The blue lines show the \nth{16}, \nth{50} and \nth{84} percentiles of the residuals. \textbf{Top right:} Same as top left, but selecting only the high-Galactic-latitude sky ($\left| b \right| > \deg{45}$). An overall offset of +0.014~mag is found. \textbf{Bottom right:} As top left and top right, but selecting instead only the Galactic plane ($\left| b \right| < \deg{5}$). A similar overall reddening offset is found as at high Galactic latitudes, but at large reddening, Bayestar17 finds more reddening than the old map, Bayestar15. This indicates that Bayestar17 extends to greater dust optical depths. \textbf{Bottom left:} Reddening residuals as a function of distance modulus, in a region of the sky where Bayestar17 finds less cumulative reddening than Bayestar15. The discrepancy is due to an increase in reddening in Bayestar15 at distance modulus $\mu \sim 11 \, \mathrm{mag}$, corresponding to a height of $\sim$400~pc above the plane of the Galaxy. This increase in reddening in Bayestar15 is likely spurious, as it occurs several dust scale heights above the plane of the Galaxy.}
    \label{fig:corr_bmk_minus_bayestar}
\end{figure*}

\begin{figure*}
    \begin{subfigure}[t]{\textwidth}
    	\includegraphics[width=\textwidth]{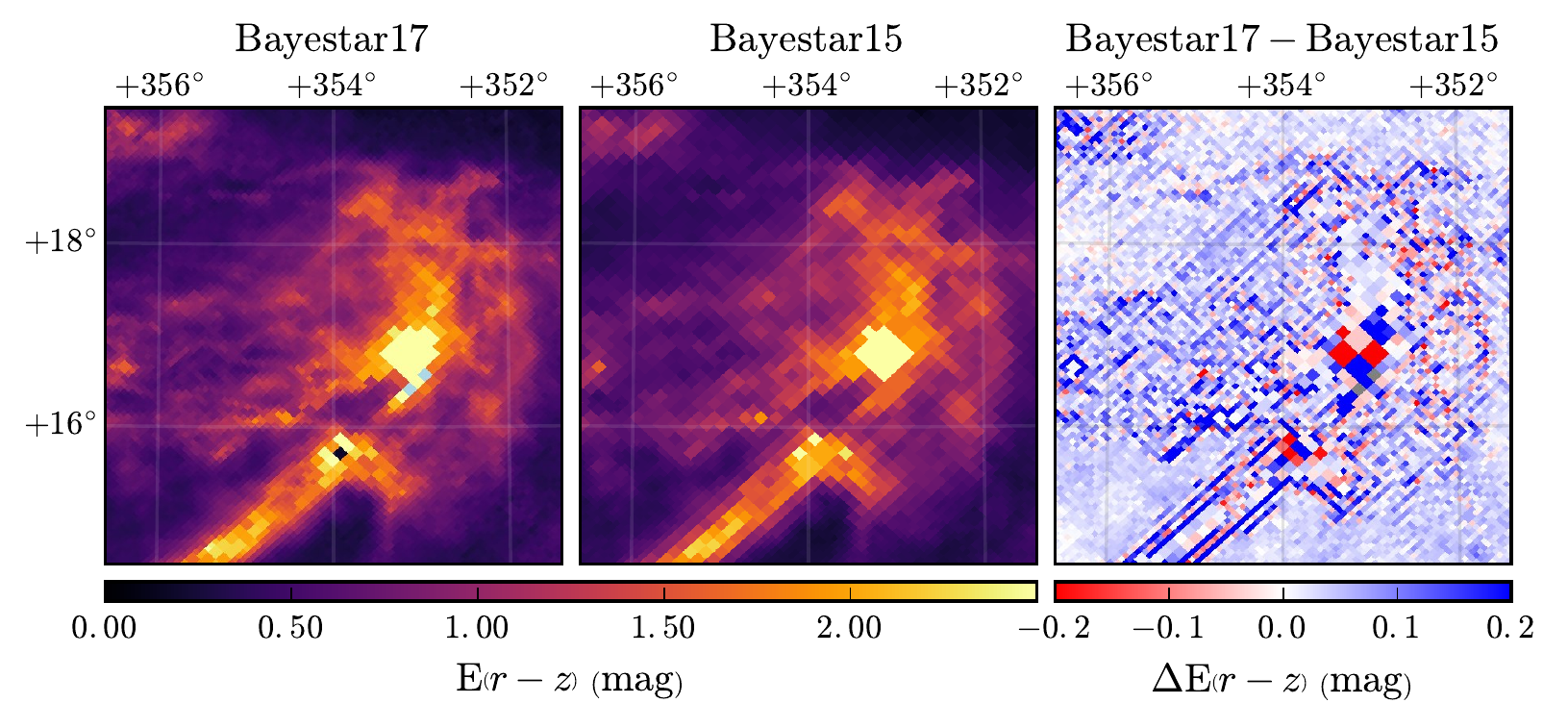}
        \subcaption{}
        \label{fig:bmk_minus_bayestar_rho_oph}
    \end{subfigure}
    \begin{subfigure}[t]{\textwidth}
    	\includegraphics[width=\textwidth]{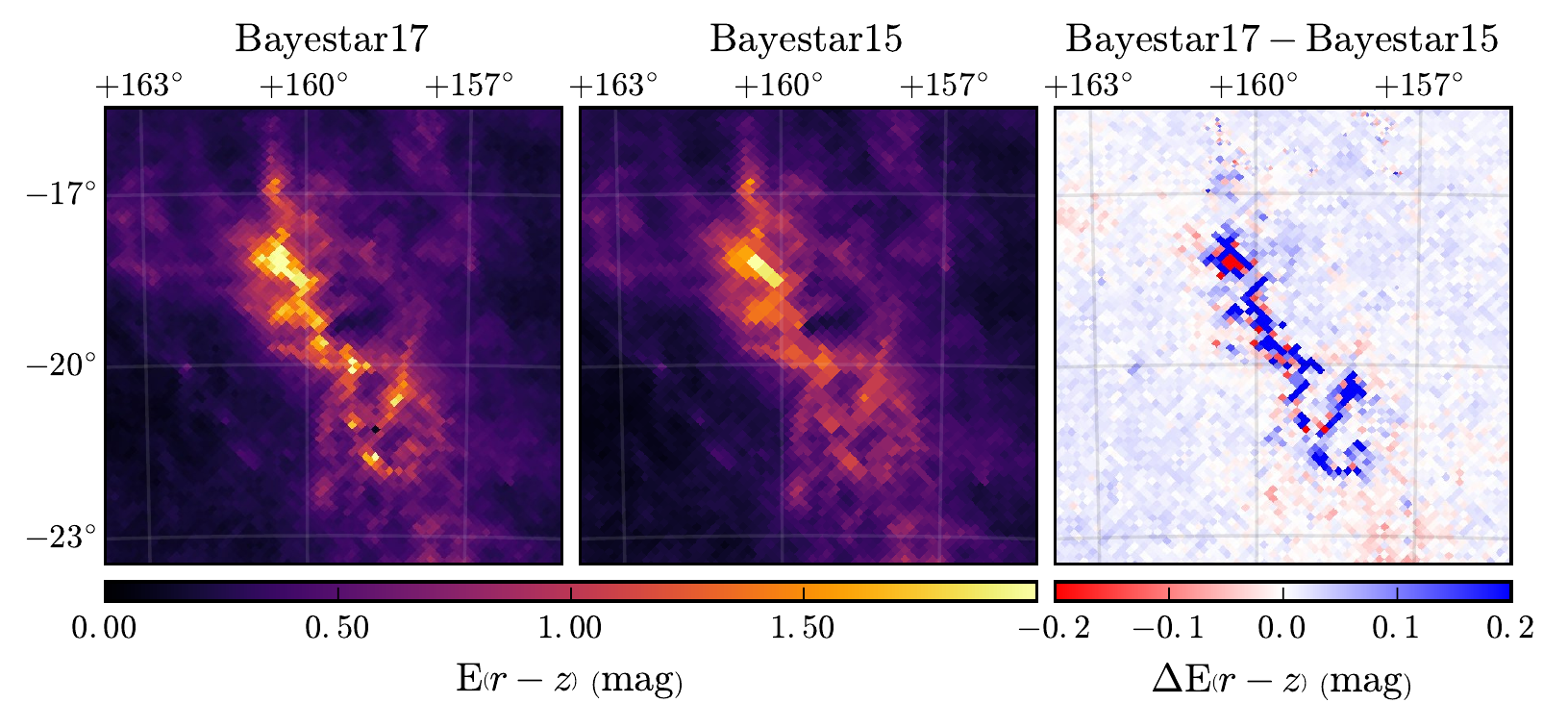}
        \subcaption{}
        \label{fig:bmk_minus_bayestar_perseus}
    \end{subfigure}
    \caption{A close-in view of the residuals between the new 3D dust map and the Bayestar15 3D dust map, in the vicinity of the \mbox{$\rho$ Ophiuchi} cloud complex (\textbf{top row}) and the Perseus cloud complex (\textbf{bottom row}). In the top row, reddening has been integrated to infinite distance, while in the bottom row, reddening has been integrated to 800~pc. As our dust maps are probabilistic, each panel shows median quantities. The new dust map (\textbf{left column}) has finer angular resolution than the Bayestar15 dust map (\textbf{center column}). The finer resolution is especially visible in top row, where sharp filamentary structures in $\rho$ Ophiuchi produce gradients in the residual map.}
    \label{fig:bmk_minus_bayestar_zoomin}
\end{figure*}

As discussed in \cref{sec:method-changes}, the major differences in the way in which our new 3D dust map was compiled, vs. the Bayestar15 dust map, are:
\begin{itemize}
    \item The use of an additional 1.5 years of PS1 photometry,
    \item The use of a new extinction vector, and
    \item Finer angular resolution in areas of increased dust opacity.
\end{itemize}
To convert the Bayestar15 map into $\Erz$, we use the $R_{V} = 3.1$ extinction vector given by \citet{Schlafly2011}, as this is the extinction vector that was assumed when constructing Bayestar15. To convert Bayestar17 into $\Erz$, we use the extinction coefficients given in \cref{tab:extinction-vector}, based on the work of S16. In \cref{fig:bmk_minus_bayestar_allsky}, we plot the difference in integrated reddening between the new and old dust maps across the entire PS1 footprint.

At high Galactic latitudes ($\left| b \right| > \deg{45}$), the new map, Bayestar17, infers $\sim$0.014~mag greater integrated $\rps-\zps$ reddening than Bayestar15, with a scatter of less than 0.015~mag (see \cref{fig:corr_sfd_vs_bmk_minus_bayestar_highlat}). Deep in the plane of the Galaxy ($\left| b \right| < \deg{5}$), Bayestar17 also has a slight offset relative to the old map (of $\sim$0.03~mag), with a scatter between the two maps of less than 0.1~mag, out to a reddening of $\Erz \approx 1.5 \, \mathrm{mag}$ (see \cref{fig:corr_sfd_vs_bmk_minus_bayestar_galplane}). The overall offset between the two maps (between $\sim$0.014~mag and $\sim$0.03~mag) may be due to a small overall offset in the color zero-point of the updated stellar templates used in Bayestar17, with the newer templates being slightly bluer than the old ones used by Bayestar15. As can be seen in \cref{fig:corr_sfd_vs_bmk_minus_bayestar_highlat}, at reddenings below $\Erz \sim 0.03 \, \mathrm{mag}$, Bayestar15 predicts less reddening than Bayestar17. This is consistent with the stellar templates used by the Bayestar15 map being slightly too red. Combined with the strictly positive prior on dust reddening, this would cause Bayestar15 to infer zero reddening when true dust reddening is below a critical threshold.

In the midplane of the Galaxy, Bayestar17 appears to extend to greater dust optical depths than the old map, Bayestar15. This is apparent in \cref{fig:corr_sfd_vs_bmk_minus_bayestar_galplane}, where at SFD reddenings beyond $\Erz \sim 2 \, \mathrm{mag}$, Bayestar17 infers more reddening than Bayestar15. This is likely because of improvements to the extinction vector, which result in a better fit to highly extinguished stars. As fewer stars are rejected at high extinctions for poor model fit, Bayestar17 is able to probe deeper into the Galactic plane. The increased depth may also be due in part to the slightly deeper coverage provided by the additional 1.5 years of PS1 photometry.

As can be seen in \cref{fig:bmk_minus_bayestar_allsky}, at about $\deg{15}$ off the plane of the Galaxy, Bayestar17 tends to infer slightly less dust than the old map. This may be due to the changes to the line-of-sight prior described in \cref{sec:method-changes}. Specifically, the new prior more strenuously requires minimal reddening far off the plane of the Galaxy. As can be seen in \cref{fig:corr_distmod_vs_bmk_minus_bayestar_redstrip}, the old map tends to infer a slight increase in reddening at a distance of $\mu \sim 11 \, \mathrm{mag}$ in this region of the sky. Because the new prior favors less dust far off the midplane of the Galaxy, Bayestar17 does not infer this increase in reddening, and therefore infers $\sim$0.02~mag less dust reddening in \Erz.

Due to our changes to the pixelization scheme and the increased number of sources in the final PS1 data release, Bayestar17 generally traces the structure of dense dust clouds with finer angular resolution. An example of this is given in \cref{fig:bmk_minus_bayestar_zoomin}, which shows the regions around the $\rho$ Ophiuchi cloud complex (top panels) and the Perseus cloud complex (bottom panels). Due to the finer angular resolution of Bayestar17, the edges of filamentary structures are clearly visible in the difference image.

\subsubsection{Comparison with Sale14}
\label{sec:iphas-comparison}

\begin{figure*}
	\includegraphics[width=\textwidth]{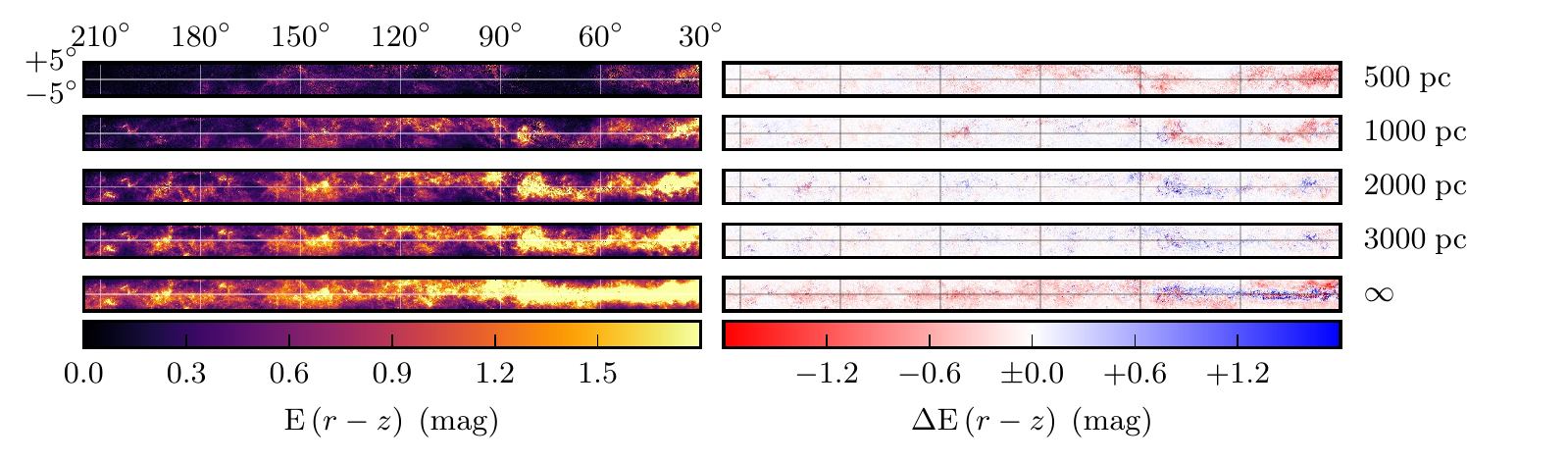}
    \caption{A comparison of Bayestar17 with Sale14, which covers the northern Galactic plane. The left panels show Bayestar17 over this footprint, integrated out to different distances. The right panel shows the difference between Bayestar17 and Sale14 out to these same distances. Positive values (blue) indicate that Bayestar17 infers greater cumulative reddening. In each pixel, a random sample of Bayestar17 and Sale14 is drawn.}
    \label{fig:bmk_minus_iphas}
\end{figure*}

\begin{figure}
	\includegraphics[width=\linewidth]{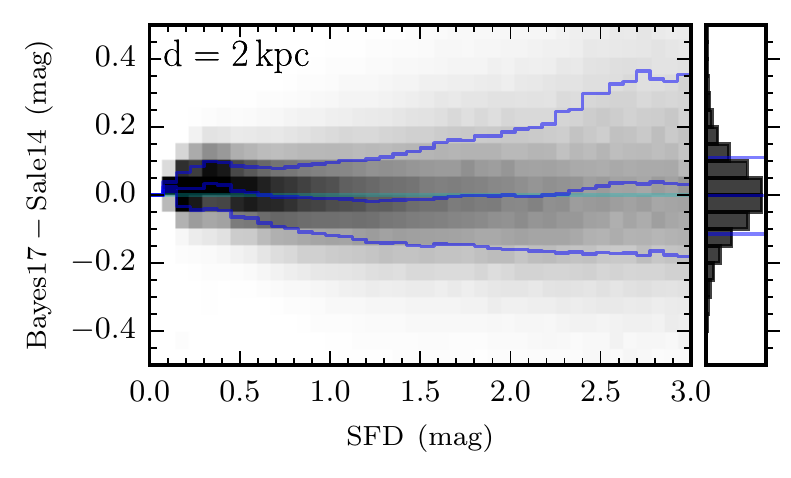}
    \caption{Comparison of Bayestar17 with Sale14, integrated out to 2~kpc. For each bin of SFD reddening, we show the histogram of reddening residuals. All units are in magnitudes of \Erz.}
    \label{fig:corr_sfd_vs_bmk_minus_iphas_2000pc}
\end{figure}

\citet{Sale2014b} maps the northern Galactic plane ($\left| b \right| < \deg{5}$, $\deg{30} < \ell < \deg{215}$) using a method that simultaneously infers stellar distances and types, as well as the line-of-sight dust distribution. The Sale14 map is based on stellar photometry from IPHAS \citep{Drew2005-IPHAS}. Sale14 reports extinction in terms of $A_0$, the extinction at 5495~\AA. In order to compare this extinction measure with \Erz, we first convert $A_0$ to $V$-band extinction, using the median relation found in \citet{Sale2014b}, namely $A_0 = 1.003 A_V$. Then, we use the $R_V = 3.1$ coefficients from Table 6 of \citet{Schlafly2011}. \cref{fig:corr_sfd_vs_bmk_minus_iphas_2000pc} shows the correlation between cumulative reddening to 2~kpc in Bayestar17 and Sale14, finding good agreement between the two maps. \cref{fig:bmk_minus_iphas} shows the difference in cumulative reddening, out to five distances, between Bayestar17 and Sale14. The two maps agree well at 2~kpc and 3~kpc, although within the nearest kiloparsec, some clouds are inferred to be closer in Sale14 than in Bayestar17. This can be seen especially in the residual map to 500~pc, where Sale14 generally infers a greater dust column than Bayestar17.

\subsubsection{Comparison with Planck14}
\label{sec:planck-comparison}

\begin{figure}
	\includegraphics[width=\linewidth]{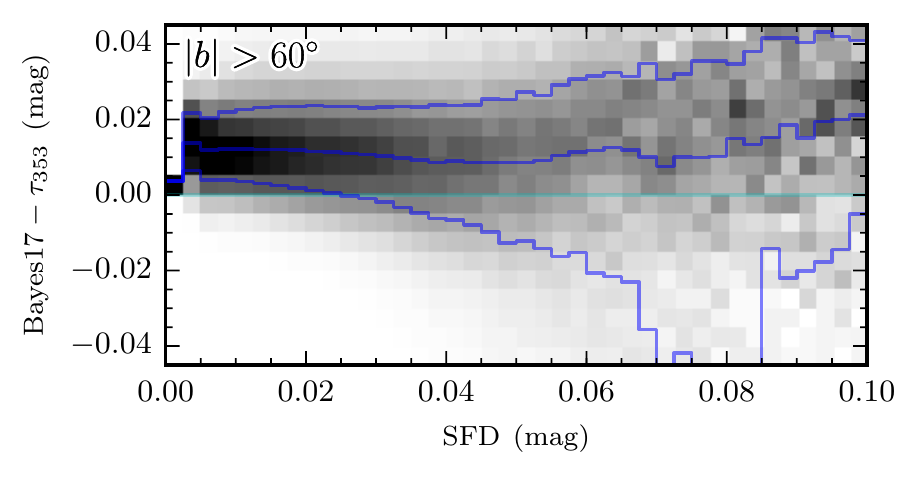}
    \caption{Comparison of Bayestar17 with the Planck14 $\tau_{353}$ dust map at high Galactic latitudes. For each bin of SFD reddening, we show the histogram of reddening residuals. All units are in magnitudes of \Erz. At these high Galactic latitudes and low reddenings, our map predicts 0.01~mag greater reddening than $\tau_{353}$, and the slope of our map vs. $\tau_{353}$ is close to unity.}
    \label{fig:corr_sfd_vs_bmk_minus_planck_tau_highlat}
\end{figure}

\begin{figure*}
	\includegraphics[width=\textwidth]{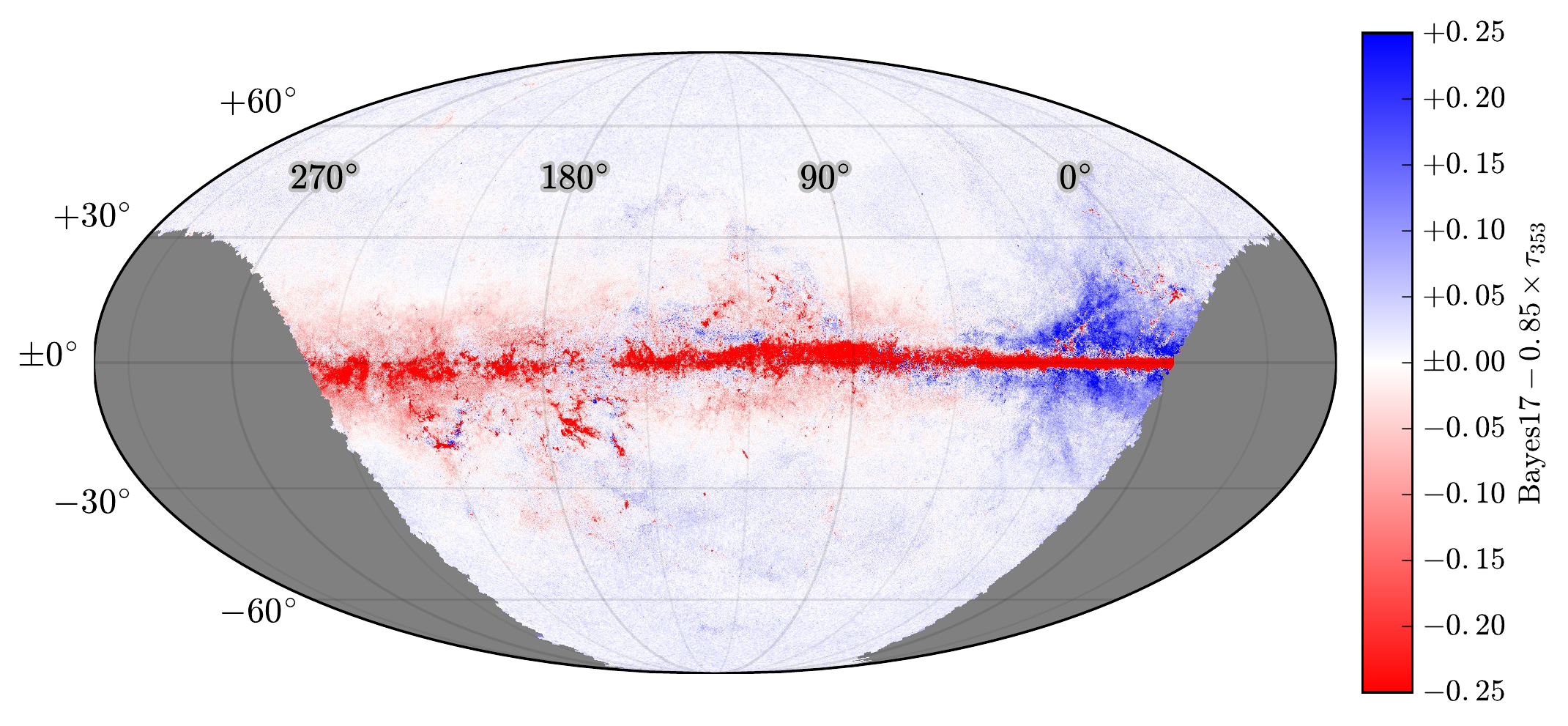}
    \caption{An all-sky map of the residuals between our new 3D dust map, Bayestar17, and the Planck14 optical-depth-based 2D dust map, in magnitudes of \Erz. In order to reduce the 3D dust map to two dimensions, it is integrated out to infinite distance. As the 3D dust map is probabilistic, we draw a random sample of cumulative dust reddening in each pixel. The two dust maps agree at high Galactic latitudes to within $\sim$15\%. Because the far infrared dust optical depth is small compared to the optical/near-infrared optical depth, the Planck dust map traces the dust to a greater column density in the plane of the Galaxy.}
    \label{fig:bmk_minus_tau_allsky}
\end{figure*}

\begin{figure*}
    \begin{subfigure}[t]{0.49\textwidth}
        \centering
        \includegraphics[width=\linewidth]{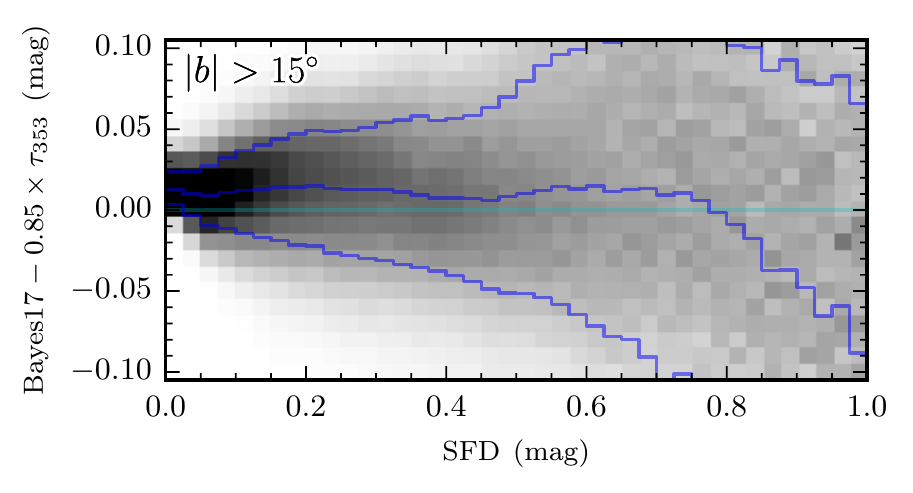}
        \subcaption{}
        \label{fig:corr_sfd_vs_bmk_minus_planck_tau}
    \end{subfigure}
    \begin{subfigure}[t]{0.49\textwidth}
        \centering
        \includegraphics[width=\linewidth]{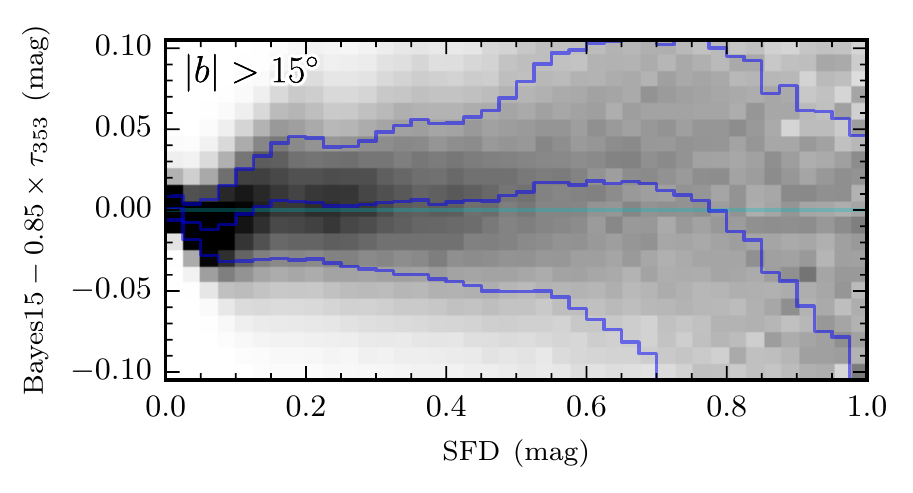}
        \subcaption{}
        \label{fig:corr_sfd_vs_bayestar_minus_planck_tau}
    \end{subfigure}

    \caption{\textbf{Left:} Reddening residuals between Bayestar17 and the Planck14 $\tau_{353}$ 2D dust map, as a function of reddening, as measured by SFD. All reddenings are in magnitudes of \Erz. For each bin of SFD reddening, we show the histogram of reddening residuals. The small panel to the right shows the overall histogram of residuals, combining all SFD reddening bins. We restrict the comparison to Galactic latitudes of $\left| b \right| > \deg{15}$. For the Bayestar17, we choose random samples of cumulative reddening. The blue lines show the \nth{16}, \nth{50} and \nth{84} percentiles of the residuals. \textbf{Right:} Reddening residuals between the Bayestar15 3D dust map and $\tau_{353}$.}
    \label{fig:corr_sfd_vs_bmk_bayestar_minus_planck_tau}
\end{figure*}

\begin{figure*}
    \centering
    \begin{subfigure}[t]{0.49\textwidth}
        \centering
    	\includegraphics[width=\linewidth]{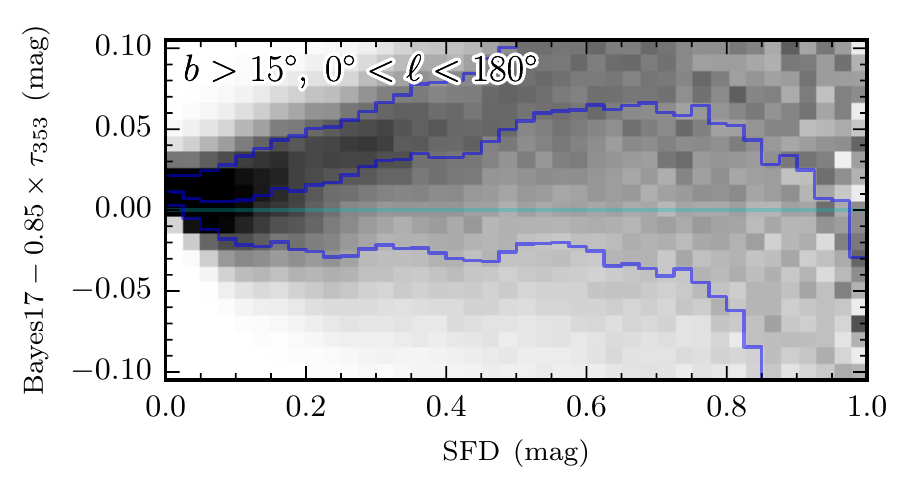}
        \phantomsubcaption{}
        \label{fig:corr_sfd_vs_bmk_minus_planck_tau_quadrant_A}
    \end{subfigure}
    \hfill
    \begin{subfigure}[t]{0.49\textwidth}
        \centering
    	\includegraphics[width=\linewidth]{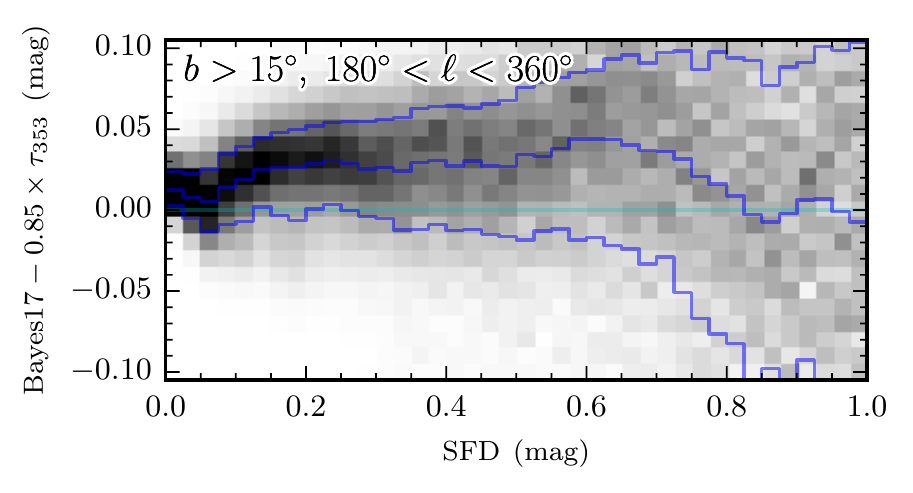}
        \phantomsubcaption{}
        \label{fig:corr_sfd_vs_bmk_minus_planck_tau_quadrant_B}
    \end{subfigure}

    \begin{subfigure}[b]{0.49\textwidth}
        \centering
    	\includegraphics[width=\linewidth]{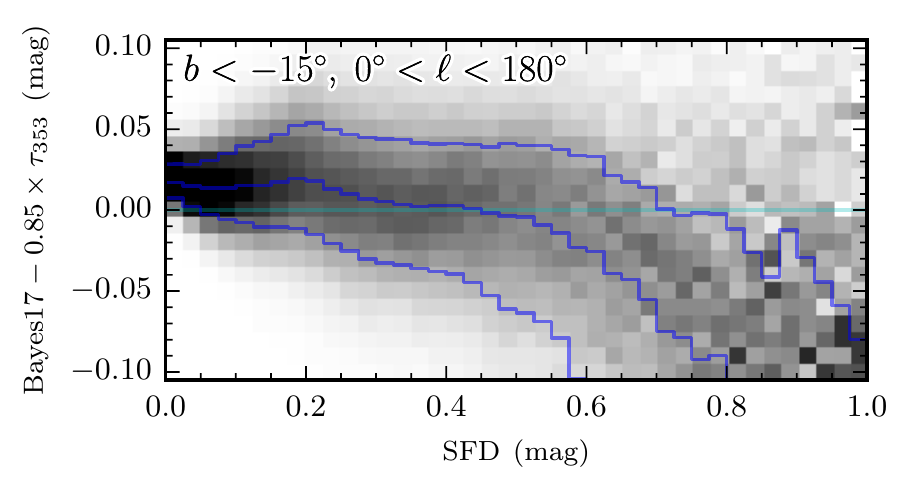}
        \phantomsubcaption{}
        \label{fig:corr_sfd_vs_bmk_minus_planck_tau_quadrant_C}
    \end{subfigure}
    \hfill
    \begin{subfigure}[b]{0.49\textwidth}
        \centering
    	\includegraphics[width=\linewidth]{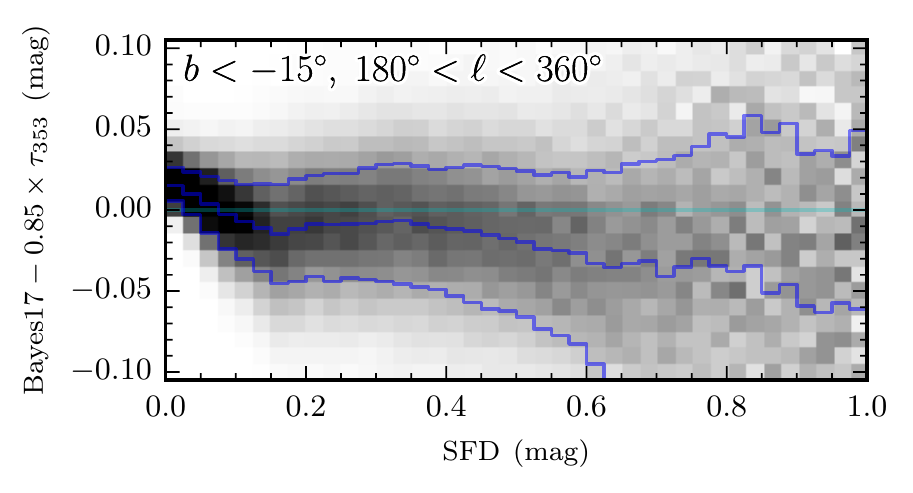}
        \phantomsubcaption{}
        \label{fig:corr_sfd_vs_bmk_minus_planck_tau_quadrant_D}
    \end{subfigure}

    \caption{Residuals between Bayestar17 and Planck14 $\tau_{353}$, in magnitudes of \Erz, as a function of SFD \Erz reddening. Each panel shows a different quarter of the sky, illustrating that the correlation between our 3D dust map and the Planck dust map varies over the sky.}
    \label{fig:corr_bmk_minus_planck_quadrants}
\end{figure*}

Planck14 fits a simple dust emission model to all-sky far-infrared Planck and IRAS data. The dust is treated as a modified blackbody, described by an optical depth, temperature and spectral index. Both the dust optical depth at 353~GHz, $\tau_{353}$, and the integrated radiance, $\mathcal{R}$, can be used to trace dust reddening. We will hereafter refer to Planck optical-depth-based reddenings as ``$\tau_{353}$.''

Planck14 reddenings are given in magnitudes of \EBV, so we convert them to \Erz using the $R_{V} = 3.1$ extinction coefficients of \citet{Schlafly2011}, leading to the relation
\begin{align}
    \Erz = 1.07 \, \EBV \, .
\end{align}
As the extinction vector used by Bayestar17, derived in \citet{Schlafly2016-extcurve}, does not contain entries for $B$ and $V$, a conversion using this extinction vector is not possible.

Planck14 gives a linear calibration between optical depth at 353~GHz and $\EBV$ derived from quasar colors in regions of the high-Galactic-latitude sky with low H\textsc{i} column density. At Galactic latitudes of $\left| b \right| > \deg{60}$, Bayestar17 agrees well with Planck14, with a constant offset of 0.01~mag in \Erz, as can be seen in \cref{fig:corr_sfd_vs_bmk_minus_planck_tau_highlat}. However, at intermediate Galactic latitudes, we find greater reddening than Planck14. This may be due to variation in the extinction curve, which would cause the ratio of dust FIR optical depth to optical/NIR reddening to vary. This interpretation is supported by the fact that we find agreement in the scale of reddening at high Galactic latitudes, where Planck14's FIR optical depth--to--reddening relation relation is calibrated, but discrepancies at lower Galactic latitudes. In the remainder of this section, we therefore compare Bayestar17 to $0.85 \times \tau_{353}$, removing the linear trend between $\tau_{353}$ and Bayestar17 seen at intermediate Galactic latitudes.

\cref{fig:bmk_minus_tau_allsky} shows the variation in dust reddening between our new 3D map, integrated to infinite distance, and $0.85 \times \tau_{353}$. In low-reddening regions of the sky, the two maps agree well, once the linear trend noted above is removed. Maps based solely on stellar photometry, as ours is, cannot trace dust beyond the farthest observable stars. The limiting distance of our map is therefore dependent on the line-of-sight extinction. Thus, in the Galactic plane, where the dust column is largest, the Planck map sees more total dust than does our 3D dust map.

One salient feature in \cref{fig:bmk_minus_tau_allsky} that deserves note is the extended region towards the Galactic center where Bayestar17 (like Bayestar15 before it) infers greater reddening than Planck14. This region is too extended to line up well with the Galactic bulge, but it does align well with nearby dust that lies within $\lesssim 500 \, \mathrm{pc}$. This suggests that nearby dust in the region $\deg{45} \lesssim \ell \lesssim \deg{0}$ has different FIR emission and/or optical/NIR extinction properties than dust elsewhere. Our method is more sensitive to changes in stellar colors, while Planck14 is more sensitive to changes in total dust column density. Therefore, either a steepening of the extinction spectrum (i.e., greater reddening per unit extinction, corresponding to lower $R_{V}$), greater FIR emissivity, or systematically overestimated dust temperatures would create the observed residuals. We see similar residuals when comparing with SFD, indicating that they are not due to any peculiar feature of Planck14.

\cref{fig:corr_sfd_vs_bmk_minus_planck_tau} shows the correlation in inferred integrated reddening between Bayestar17 and the Planck14 $\tau_{353}$ dust map. Out of the plane of the Galaxy, the two maps agree to within $\sim$10\% after $\tau_{353}$ is rescaled by 0.85. However, the exact level of agreement varies over the sky, as shown in \cref{fig:corr_bmk_minus_planck_quadrants}, which shows the correlation between our 3D map and the Planck optical-depth-based 2D dust map in four regions of the sky. As can be seen in \cref{fig:corr_bmk_minus_planck_quadrants}, the slope of the rescaling which would be required to bring $\tau_{353}$ into alignment with Bayestar17 is different from one region of the sky to another, which could be indicative of systematic changes in the extinction curve (e.g., varying $R_{V}$). Such variations are not implausible, given the large-scale variations in $R_{V}$ found by \citet{Schlafly2016-RV-3D}.

\cref{fig:corr_sfd_vs_bayestar_minus_planck_tau} shows the correlation in integrated reddening between the old 3D dust map, Bayestar15, and the Planck14 dust map. At low reddening, Bayestar15 infers less reddening than Planck out to a reddening of $\sim$0.05~mag in \Erz. Bayestar15 tends to infer zero reddening when the true reddening is below a few hundredths of a magnitude. This is again consistent with the stellar templates used in Bayestar15 being slightly too red, as discussed in \cref{sec:comparison-with-bayestar}. This effect is not seen when comparing the new 3D dust map, Bayestar17, with the Planck14 dust map (see \cref{fig:corr_sfd_vs_bmk_minus_planck_tau}), indicating that this low-reddening systematic effect is largely resolved in Bayestar17.

\section{Discussion}
\label{sec:discussion}

This map is an update on the 3D map released by Bayestar15. The advantages of our new map, Bayestar17, are that it includes an additional year-and-a-half of PS1 data, a more accurate extinction vector \citep{Schlafly2016-extcurve}, and finer angular resolution in heavily extinguished regions of the Galaxy. Bayestar17 recovers largely the same qualitative features as the previous map, Bayestar15, but probes to slightly greater depths. With distance projected out, Bayestar17 correlates better, especially at low reddenings, with the 2D Planck FIR optical-depth-based dust map.

For extra-Galactic astronomy, where only the integrated extinction or reddening is typically of interest, dust maps that are based directly on stellar photometry provide a valuable complement to FIR emission-based maps. The two methods have different types of systematics. Emission-based maps depend on a detailed modeling of the temperature, optical depth, and the shape of the emission spectrum -- the latter possibly based on a physical model of dust properties, as in \citet{PlanckDustPhysical2016}, or on a simple empirical model, as in SFD, Planck14 and \citet{Meisner2015-Planck2Component}. By contrast, dust maps based on stellar photometry more directly measure reddening, but depend on accurate modeling of stellar spectral energy distributions and assumptions about the shape of the dust extinction spectrum. As shown in \cref{sec:planck-comparison}, at low extinctions (where star-based maps can still observe background stars), these two methods generally have a scatter of $\sim$10\%, although the slope of the relation between the two types of maps can vary by $\sim$15\% across different regions of the sky. Variations in residuals across the sky between these two types of dust map are potentially an indication of variation in the relation between the extinction and emission properties of dust. Such variations are expected, as dust grain composition and size are known to vary (for an overview of interstellar dust properties, see \citealt{Draine2003}), although the large spatial extent of these variations is surprising. S16 measures one side of this variation -- the variation in the dust extinction vector -- by measuring the variation in the colors of stars of known spectral types. \citet{Schlafly2016-RV-3D} combines $R_{V}$ measurements for individual stars with the three-dimensional distribution of dust (from Bayestar15) to map variation in dust $R_{V}$ in three dimensions across the Galaxy. Unmodeled variation in the dust extinction vector is one source of residuals between our 3D dust map and emission-based dust maps, but incompletely modeled variation in dust emission properties must also contribute to the residuals. In addition, FIR zodiacal light (from dust grains in the Solar system) and contamination from large-scale structure (due to dust emission in distant galaxies) introduce systematics into far-infrared emission-based dust maps, which are not present in maps based on stellar colors.

Our 3D dust map currently assumes a universal dust extinction law (see \cref{sec:method-summary}) with $R_{V} \approx 3.1$. Allowing the dust extinction spectrum to vary is a priority for constructing a more accurate 3D dust map. One way of doing this would be to infer not only a dust density in each voxel, but also a dust $R_{V}$. In order to avoid producing an underconstrained model, one should take advantage of spatial smoothness in both dust density and $R_{V}$, imposing a prior that correlates dust properties in nearby voxels. A number of methods have been proposed to impose a smoothness condition on the 3D dust density field. \citet{Sale2014a-GPs} and \citet{Rezaei2016} lay out differing methods of imposing smoothness on a 3D dust map by treating the dust density field as a Gaussian process.

It is also possible to impose smoothness on the dust density field as a post-processing step on a map constructed from independent sightlines, assuming that a sufficient number of samples have been saved for each sightline. A selection of one sample from each sightline may be regarded as a sample from the entire map. By applying a new, joint prior linking nearby sightlines, one may reweight samples of the full map. This process produces a set of realizations of the full map in which neighboring sightlines are consistent with one another, in that dust density varies smoothly with angle. For a map with $n_{\mathrm{pix}}$ angular pixels and $n_{\mathrm{samples}}$ stored samples of the distance--reddening relation in each angular pixel, the total number of realizations of the full map is given by
\begin{align}
    n_{\mathrm{realizations}} = \left( n_{\mathrm{samples}} \right)^{n_{\mathrm{pix}}} \, .
\end{align}
For our map, $n_{\mathrm{pix}} = 3.42 \times 10^{6}$ and $n_{\mathrm{samples}} = 500$, yielding $n_{\mathrm{realizations}} \approx 10^{10^{7}}$. Clearly, reweighting the posterior probability of every realization of the full map is not feasible. However, this problem lends itself to the types of strategies commonly pursued to sample from the 2D Ising model of ferromagnetism. Such strategies can yield a representative set of realizations of the full map. We leave more detailed discussion of this method to future work.

Incorporating new types of datasets will also improve the accuracy of 3D dust maps. In particular, parallaxes measured by the ESA space telescope Gaia \citep{Lindegren1994} will provide future maps with tight priors on the distances to a billion stars, effectively removing distance as a fitting parameter for a large subset of stars. Distance determinations and stellar types provided by Gaia will also aid in the construction of more accurate stellar photometric templates. Finally, Gaia's spectrophotometry can be used to infer stellar type and reddening, in much the same way as we currently use PS1 and 2MASS photometry.

We intend to extend 3D reddening mapping to cover the entire sky, and to that end, we have imaged the portion of the Galactic plane visible only from the Southern Hemisphere with the Dark Energy Camera (DECam). The resulting 5-band optical/NIR survey will be described in an upcoming paper (Schlafly et al., \textit{in prep.}). As various other projects fill out DECam coverage of the remainder of the Southern sky, a high-quality, $4 \pi$-steradian, 3D map of Galactic reddening will become possible.

\section{Conclusions}
\label{sec:conclusions}

We have produced a new 3D map of interstellar dust reddening, based on stellar photometry from PS1 and 2MASS. The new map, Bayestar17, covers the $3 \pi$ steradians north of declination $\delta = -\deg{30}$. It is compiled using a nearly identical method to Bayestar15, but with the addition of an extra 1.5 years of PS1 photometry, an updated extinction vector from S16, and finer angular resolution in regions of high extinction. Bayestar17 probes somewhat deeper than the old map, Bayestar15, in the plane of the Milky Way. In low-extinction regions of the sky, Bayestar17 shows better agreement with 2D FIR emission-based dust maps, such as Planck14. Above the plane of the Galaxy, the average scatter between Bayestar17 and Planck14 is $\sim$10\%, although the slope between the two maps varies across the sky by $\sim$15\%.

Three dimensional dust maps based on stellar photometry have a number of advantages over FIR dust emission-based maps, including the ability to trace dust in 3D, and the fact that they more directly trace the quantity of interest -- dust extinction. Maps based on stellar photometry will not only be of use in tracing extinction in the plane of the Galaxy, where the distance-dependence of extinction is crucial, but can also provide a complementary measurement of dust reddening for extra-Galactic astronomy. Since the systematic uncertainties in stellar photometric reddenings are very different from the types of systematic effects in emission-based maps, one can expect the two types of measurements to have uncorrelated errors. In particular, stellar photometric reddenings do not suffer from contamination from zodiacal light or large-scale structure. Maps based on stellar photometry also have the potential to trace changes in the extinction curve (see, for example, \citealt{Schlafly2016-RV-3D}). Producing a 3D dust map that traces spatial variation in both dust density and the extinction curve, taking into account the smoothness of these quantities in space, will be an important development in dust mapping efforts.

Our new 3D dust map can be accessed at \url{http://argonaut.skymaps.info}, or by its digital object identifier (DOI): \href{http://dx.doi.org/10.7910/DVN/LCYHJG}{10.7910/DVN/LCYHJG}. In addition, the Python package \texttt{dustmaps} provides an interface for Bayestar17, Bayestar15, SFD and other three- and two-dimensional dust maps.

\section*{Acknowledgements}
The Pan-STARRS1 Surveys (PS1) have been made possible through contributions of the Institute for Astronomy, the University of Hawaii, the Pan-STARRS Project Office, the Max-Planck Society and its participating institutes, the Max Planck Institute for Astronomy, Heidelberg, and the Max Planck Institute for Extraterrestrial Physics, Garching, The Johns Hopkins University, Durham University, the University of Edinburgh, Queen's University Belfast, the Harvard-Smithsonian Center for Astrophysics, the Las Cumbres Observatory Global Telescope Network Incorporated, the National Central University of Taiwan, the Space Telescope Science Institute, the National Aeronautics and Space Administration under Grant No. NNX08AR22G issued through the Planetary Science Division of the NASA Science Mission Directorate, the National Science Foundation under grant No. AST-1238877, the University of Maryland, and E\"{o}tv\"{o}s Lor\'{a}nd University (ELTE). This publication makes use of data products from the 2MASS, which is a joint project of the University of Massachusetts and the Infrared Processing and Analysis Center/California Institute of Technology, funded by the National Aeronautics and Space Administration and the National Science Foundation. The computations in this paper were run on the Odyssey cluster supported by the FAS Science Division Research Computing Group at Harvard University. G.G. and D.F. are supported by NSF grant AST-1312891. E.S. acknowledges support for this work provided by NASA through Hubble Fellowship grant HST-HF2-51367.001-A awarded by the Space Telescope Science Institute, which is operated by the Association of Universities for Research in Astronomy, Inc., for NASA, under contract NAS 5-26555.




\bibliographystyle{mnras}
\bibliography{arxiv}




%
%


\bsp	
\label{lastpage}
\end{document}